\documentclass[%
reprint,
 amsmath,
 amssymb,
 aps,
 prl,
]{revtex4-1}

\usepackage{xcolor}
\usepackage[colorlinks=true,citecolor=blue]{hyperref}
\usepackage{graphicx}
\usepackage{subfigure}
\usepackage{bookmark}
\usepackage{hyperref}
\usepackage[normalem]{ulem}
\usepackage{dsfont} 
\usepackage{upgreek}
\usepackage{booktabs}

\newcommand{\comma}{~,}
\newcommand{\fullstop}{~.}

\newcommand{\hc}{\mathrm{h.c.}}
\newcommand{\abs}[1]{\left\vert #1 \right\vert}
\newcommand{\cabs}[1]{\vert #1 \vert}
\newcommand{\erw}[1]{\left\langle #1 \right\rangle}
\newcommand{\cerw}[1]{\langle #1 \rangle}
\newcommand{\komm}[2]{\left[ #1, #2 \right]}

\newcommand{\cakomm}[2]{\lbrace #1, #2 \rbrace}
\newcommand{\ket}[1]{\left\vert #1 \right\rangle}
\newcommand{\bra}[1]{\left\langle #1 \right\vert}

\renewcommand{\d}{\mathrm{d}}

\newcommand{\omopt}{\omega_\mathrm{o}}
\newcommand{\ommech}{\omega_\mathrm{m}}
\newcommand{\omspin}{\omega_\mathrm{s}}
\newcommand{\ompump}{\omega_\mathrm{pump}}
\newcommand{\ompr}{\omega_\mathrm{pr}}
\newcommand{\omprrot}{\tilde{\omega}_\mathrm{pr}}

\newcommand{\gom}{g_0}
\newcommand{\gspin}{g_\mathrm{sm}}
\newcommand{\nth}{n_\mathrm{th}}
\newcommand{\Gmech}{\Gamma_\mathrm{mech}}
\newcommand{\Deltaopt}{\Delta_\mathrm{opt}}
\newcommand{\Deltaoptbar}{\bar{\Delta}_\mathrm{opt}}
\newcommand{\Deltaspin}{\Delta_\mathrm{sm}}
\newcommand{\Com}{C_\mathrm{om}}

\newcommand{\SNR}{\mathrm{SNR}}
\newcommand{\I}{\mathcal{I}}
\newcommand{\tmeas}{\tau_\mathrm{meas}}
\newcommand{\nmech}{n_\mathrm{mech}}
\newcommand{\ncav}{n_\mathrm{cav}}
\newcommand{\ncrit}{\nmech^\mathrm{crit}}
\newcommand{\optmode}{\hat{a}}
\newcommand{\mechmode}{\hat{b}}
\newcommand{\optflucsymbol}{d}
\newcommand{\mechflucsymbol}{c}
\newcommand{\optfluc}{\hat{\optflucsymbol}}
\newcommand{\mechfluc}{\hat{\mechflucsymbol}}
\newcommand{\1}{\mathds{1}}
\newcommand{\nmeas}{n_\mathrm{add}}
\newcommand{\ndet}{n_\mathrm{det}}

\newcommand{\xcavout}{\hat{x}_\mathrm{R,out}^\mathrm{cav}}
\newcommand{\pcavout}{\hat{p}_\mathrm{R,out}^\mathrm{cav}}
\newcommand{\xmech}{\hat{x}_\mathrm{R}^\mathrm{mech}}
\newcommand{\pmech}{\hat{p}_\mathrm{R}^\mathrm{mech}}

\newcommand{\thetitle}{Single-Spin Readout and Quantum Sensing using Optomechanically Induced Transparency}
\newcommand{\theauthors}{Martin Koppenh\"ofer$^{1}$, Carl Padgett$^{2}$, Jeffrey V.\ Cady$^{2,3}$,\\Viraj Dharod$^{2}$, Hyunseok Oh$^{2}$, Ania C.\ Bleszynski Jayich$^{2}$, and A.\ A.\ Clerk$^1$}
\newcommand{\theaffiliations}{$^1$Pritzker School of Molecular Engineering, University of Chicago, Chicago, IL 60637, USA\\$^2$Department of Physics, University of California Santa Barbara, Santa Barbara, CA 93106, USA\\$^3$Systems and Processes Engineering Corporation, Austin, TX 78737, USA}

\newcommand{\prlsection}[1]{\paragraph*{#1---}}

\def\ValueGammaMech{200\,\mathrm{kHz}}
\def\Valuekappa{2\,\mathrm{GHz}}

\def\Valueommech{6\,\mathrm{GHz}}
\def\Valuechi{27\,\mathrm{kHz}}
\def\ValuechioverGammaMech{0.13}
\def\Valuegom{200\,\mathrm{kHz}}
\def\Valuegsm{2\,\mathrm{MHz}}
\def\ValueDeltasm{150\,\mathrm{MHz}}
\def\ValueTauMeas{3.31\,\upmu\mathrm{s}}

\def\ValueTauPurcell{28\,\mathrm{ms}}
\def\ValueRatioTauPurcellOverTauMeas{8500}
\def\ValueRatiokappaoverGamma{10000}
\def\ValueDeltasmOverGamma{750}
\def\Valueapr{20.0}
\def\Valuenmech{ 137}
\def\Valuenmechcrit{5700}

\def\ValuelambdaSO{46\,\mathrm{GHz}}
\def\ValuegammaL{1.4\,\mathrm{GHz}/\mathrm{T}}
\def\ValuegammaS{14\,\mathrm{GHz}/\mathrm{T}}
\def\Valueomegamech{7.64\,\mathrm{GHz}}
\def\ValueepsilonEgx{-7.92\,\mathrm{MHz}}
\def\ValueepsilonEgy{0.0\,\mathrm{MHz}}

\begin{document}

\title{\thetitle}
\author{\theauthors}
\affiliation{\theaffiliations}
\date{\today}

\begin{abstract}
Solid-state spin defects are promising quantum sensors for a large variety of sensing targets.
Some of these defects couple appreciably to strain in the host material. 
We propose to use this strain coupling for mechanically-mediated dispersive single-shot spin readout by an optomechanically-induced transparency measurement.
Surprisingly, the estimated measurement times for negatively-charged silicon-vacancy defects in diamond are an order of magnitude shorter than those for single-shot optical fluorescence readout. 
Our scheme can also be used for general parameter-estimation metrology and offers a higher sensitivity than conventional schemes using continuous position detection. 
\end{abstract}

\maketitle


\prlsection{Introduction}

Solid-state defect spins are promising candidates to build powerful quantum sensors \cite{Acosta2009,Steinert2010,Pham2011,Wolf2015} as well as memories and repeaters for quantum communication \cite{Awschalom2018}. 
They have a small footprint \cite{Mamin2013,Staudacher2013}, straightforward operation, and are susceptible to a large variety of sensing targets, such as magnetic \cite{Taylor2008,Rondin2014} and electric fields \cite{Dolde2011} as well as temperature \cite{Acosta2010}. 
Quantum applications (e.g., entanglement-assisted metrology \cite{Giovannetti2004,Degen2017,Pezze2018}) require high-fidelity single-shot spin readout.
Optical spin readout is desirable but, unfortunately, not provided by all types of spin defects. 
Moreover, even many optically addressable spin defects fail to reach robust high-fidelity single-shot readout \cite{Awschalom2018,Barry2020}, e.g., because of low photon collection efficiencies, inconvenient optical frequencies, or limited readout times due to non-spin-conserving transitions between orbital ground and excited states.

\begin{figure}[t]
	\centering
	\includegraphics[width=0.48\textwidth]{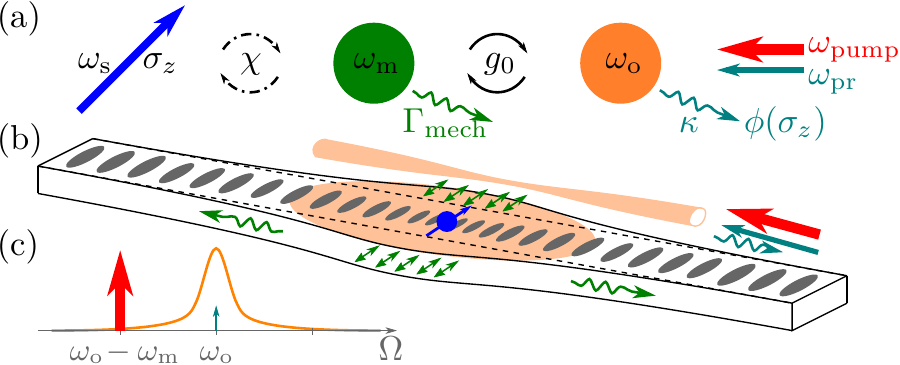}
	\caption{
		Dispersive spin readout using optomechanically-induced transparency (OMIT). 
		(a) Sketch of the considered hybrid optomechanical system. 
		A mechanical mode (green circle, center) interacts both with a single spin (blue arrow, left) via a strain-coupling-mediated dispersive interaction, and with an optical mode (orange circle, right) via optomechanical interaction. 
		The optical mode is driven by a pump and a probe laser implementing an OMIT scheme. 
		The $\sigma_z$ projection of the spin state is encoded in the phase $\phi(\sigma_z)$ of the reflected probe light. 
		All other parameters are defined in the main text. 
		(b) Sketch of a possible experimental implementation using a diamond optomechanical crystal (OMC) with an embedded spin defect (blue) strain-coupled to a mechanical breathing mode (green straight arrows). 
		The optical mode (orange) of the OMC is evanescently coupled to a tapered fiber for optical driving and homodyne detection.
		(c) Frequencies of the pump and probe lasers. 
		The solid orange curve is the Lorentzian cavity response with width $\kappa < \ommech$. 
	}
	\label{fig:sketch}
\end{figure}

These issues motivate asking whether other interactions could be harnessed for readout. 
Recently, it has been shown that some spin defects have an appreciable coupling to strain arising from mechanical vibrations in their host material  \cite{Meesala2016,Lee2017,Meesala2018}.
It has been suggested to use this strain coupling for mechanical cooling \cite{Kepesidis2016}, mechanical control of the spin defect \cite{MacQuarrie2013,MacQuarrie2015,Barfuss2015,Lee2016,Golter2016,Meesala2016}, and reservoir engineering \cite{Groszkowski2022,Kitzman2022}.
The mechanical mode can also be strongly coupled to electromagnetic modes, e.g., by shaping the host material into an optomechanical crystal (OMC) \cite{Eichenfield2009}, which enables optical control and fiber-coupled telecom-wavelength optical access, instead of more challenging free-space optical access that is often in the visible range.
Diamond OMCs with large optomechanical coupling and integrated nitrogen-vacancy (NV) defects have already been demonstrated experimentally \cite{Golter2016,Burek2016,Cady2019}.

In this Letter, we show that strain coupling can be used for another crucial functionality: it can enable rapid all-optical dispersive readout of a single solid-state spin, without any orbital excitation.   
Dispersive readout enables fast, high-fidelity, and quantum-nondemolition (QND) detection in a variety of platforms, including superconducting qubits \cite{Blais2021}, where the state of the qubit shifts the resonance frequency of a driven microwave cavity and is encoded in the phase of the microwave output field.
Using strain coupling, one could try to replicate this by replacing the microwave cavity with a driven, dispersively-coupled mechanical mode. 
Qubit readout would then require an effective homodyne detection of emitted phonons, which could be done optically using mechanics-to-optics transduction.    
The scheme we introduce mimics this kind of measurement in a simple and resource-efficient fashion, by exploiting one of the most ubiquitous effects in optomechanics:  optomechanically-induced transparency (OMIT) \cite{Schliesser2009,Agarwal2010,Weis2010,SavafiNaeini2011}, where a mechanical mode alters the density-of-states of an optical cavity. 
While OMIT has been used extensively for device calibration, we show here that, surprisingly, it also paves a powerful route to all-optical single-shot solid-state spin readout (no explicit mechanical driving or readout is needed).
Note that our OMIT-based scheme is distinct from the recently demonstrated optical readout of a superconducting qubit using optomechanical microwave-to-optical transduction \cite{Delaney2022,FN1}.

As a promising experimental example, we analyze readout of a  silicon-vacancy (SiV) defect coupled to a diamond OMC.
Surprisingly, the estimated spin readout times for realistic experimental parameters \cite{Burek2016,Cady2019,Shandilya2021,SM} are more than a factor of four shorter than the ones for optical cavity-based SiV readout \cite{Nguyen2019}, and an order of magnitude shorter than the best optical fluorescence readout times for SiV centers \cite{Sukachev2017} (which are limited by the repolarization timescale of the spin defect into its ground state and require precise alignment of the magnetic field along the SiV axis).
In contrast, our dispersive readout is in principle a QND measurement.
We stress that our protocol can be applied to other spin-defects (beyond SiV centers) with sufficiently large strain coupling but potentially no optical addressability, since we only assume coupling of an effective two-level system to a mechanical mode.

We also demonstrate that our OMIT-based sensing protocol has applications beyond qubit readout:  it can be used for parameter sensing in any optomechanical system where the mechanical frequency depends on an unknown parameter. 
It exceeds fundamental sensitivity limits that constrain standard schemes employing continuous mechanical position detection [e.g., as used in atomic-force-microscopy (AFM) \cite{Martin1987,Albrecht1991} and mass sensing \cite{Ekinci2004}].

\prlsection{The system}

We consider a standard optomechanical (OM) system, sketched in Figs.~\ref{fig:sketch}(a,b), with Hamiltonian $\hat{H}_\mathrm{om} = \omopt \optmode^\dagger \optmode + \ommech \mechmode^\dagger \mechmode - g_0 \optmode^\dagger\optmode(\mechmode + \mechmode^\dagger)$.
Here, $\optmode$ ($\mechmode$) is the annihilation operator of the optical (mechanical) mode with frequency $\omopt$ ($\ommech$), $g_0$ is the bare OM coupling strength, and $\hbar = 1$. 
Both modes interact with dissipative Markovian environments which lead to a decay of optical (mechanical) excitations at a rate $\kappa$ ($\Gmech$), with $\kappa \gg \Gmech$.
For simplicity, we envisage a several-$\mathrm{GHz}$ mechanical mode in a dilution refrigerator, such that thermal occupation is negligible \cite{FN2}.

The mechanical mode is dispersively coupled to a spin, $\hat{H}_\mathrm{sm} = \omspin \hat{\sigma}_z/2 - \chi \hat{\sigma}_z \mechmode^\dagger \mechmode$, where $\hat{\sigma}_z$ is the Pauli $z$ matrix (and commutes with the spin-only Hamiltonian), $\omspin$ is the splitting between the two energy levels, and $\chi$ is the dispersive coupling strength.  
Depending on the $\sigma_z$ projection of the spin state, the mechanical frequency is shifted by $\varepsilon = - \sigma_z \chi$. 
In principle, the mechanical frequency shift $\varepsilon$ can be measured by driving the mechanical mode with a linear drive and by measuring the phase of the phonons emitted from the mechanical mode into the substrate; this would be a mechanical analogue of a standard cavity QED dispersive readout \cite{Blais2021}. 
Of course, directly measuring these emitted phonons is infeasible in most setups.

To overcome this issue, we propose an OMIT measurement \cite{Schliesser2009,Agarwal2010,Weis2010,SavafiNaeini2011} with two laser drives, as shown in Fig.~\ref{fig:sketch}(c).
The strong red-detuned pump laser, $\ompump = \omopt - \ommech$, causes additional mechanical damping and converts part of the dissipated phonons into an optical output field, thus rendering them accessible to conventional optical homodyne detection. 
Via the OM interaction, it also converts the weak optical probe laser into a linear mechanical drive. 
Together, these enable all-optical readout of $\varepsilon$, as we now show.

Consider first the situation with only the strong pump laser.  
It allows us to separate the cavity field into a semiclassical amplitude $a \gg 1$ and quantum fluctuations $\optfluc$ around it, $\optmode = e^{-i \ompump t} (a + \optfluc)$.
Similarly, we decompose the mechanical mode $\mechmode = b + \mechfluc$ and 
linearize the OM interaction \cite{Aspelmeyer2014RMP}.
We further assume the good cavity limit $\ommech \gg \kappa$, allowing us to make a rotating wave approximation on the OM interaction.  
In a frame rotating at $\ompump$, the approximate linearized Hamiltonian is:
\begin{align}
	\hat{H} \approx \ommech \optfluc^\dagger \optfluc + (\ommech + \varepsilon) \mechfluc^\dagger \mechfluc - G \left( \mechfluc^\dagger \optfluc + \optfluc^\dagger \mechfluc \right) \comma
	\label{eqn:LinearizedHamiltonian}
\end{align}
where $G = \gom a$ is the optically-enhanced coupling strength.
At time $t=0$ the weak probe laser at frequency $\ompr$ is switched on. 
We account for this through the cavity input field, $\optfluc_\mathrm{in}(t \geq 0) = a_\mathrm{pr,in} e^{-i \ommech t} + \hat{\xi}_\mathrm{in}(t)$, where $\hat{\xi}_\mathrm{in}(t)$ is input vacuum noise and $\abs{a_\mathrm{pr,in}}^2$ is the photon flux of the probe laser.
Note that $\optfluc_\mathrm{in}$ describes a probe laser which is resonant with the optical cavity in the lab frame, cf.\ Fig.~\ref{fig:sketch}(c).
We also considered a detuned probe laser but found the resonant case to be optimal for qubit readout \cite{SM}.

\prlsection{Signal-to-noise ratio (SNR)}

For $a_\mathrm{pr,in} \geq 0$ and $\kappa \gg \Gmech$, the mechanical frequency shift $\varepsilon$ is encoded in the $\varphi = \pi/2$ quadrature of the optical output field $\optfluc_\mathrm{out}(t) = \sqrt{\kappa} \optfluc(t) + \optfluc_\mathrm{in}(t)$ and can be measured by optical homodyne detection.
Using the measurement operator describing the integrated homodyne current from $t=0$ to $t = \tau$,
\begin{align}
	\hat{\I}(\tau) = \sqrt{\kappa} \int_0^\tau \d t\, \left[ e^{i \varphi} e^{-i \ommech t} \optfluc_\mathrm{out}^\dagger(t) + \hc \right] \comma
	\label{eqn:Itau}
\end{align}
the SNR at time $\tau$ of our qubit $\sigma_z$ measurement is defined as \cite{Didier2015}
\begin{align}
	\SNR^2(\tau) 
	= \frac{\cabs{ \cerw{\hat{\I}(\tau)}_{-\chi} - \cerw{\hat{\I}(\tau)}_{+\chi}}^2}{\cerw{[\widehat{\delta \I}(\tau)]^2}_{-\chi} + \cerw{[\widehat{\delta \I}(\tau)]^2}_{+\chi}} \comma
	\label{eqn:DefinitionSNR}
\end{align}
where $\widehat{\delta\I}(\tau) = \hat{\I}(t) - \cerw{\hat{\I}(\tau)}_\varepsilon$ and $\cerw{\cdot}_\varepsilon$ denotes an expectation value with the mechanical resonance frequency shifted by $\varepsilon$. 
We focus on the usual limit $G \ll \kappa$ where there is no many-photon OM strong coupling, and where $\chi \ll \kappa$.  
Note that the effects of $\chi$ can still be non-perturbative if $\chi \gtrsim \Gmech$. 
Using the Heisenberg-Langevin equations for our system \cite{SM}, we find
\begin{align}
	\SNR^2(\tau) &= 8 \abs{a_\mathrm{pr,in}}^2 \left( \frac{\Com}{1 + \Com} \right)^2 \sin^2(2 \xi) \tau \left[ 1 - F(\tau) \right]^2 \comma
	\label{eqn:SNR}
\end{align}
where $F(\tau) = \frac{1}{\chi \tau} [\sin(2 \xi) - \sin(2 \xi + \chi \tau) e^{- \Gmech (1 + \Com) \tau/2}]$. 
The OM cooperativity $\Com = 4 G^2/\kappa\Gmech$ can be tuned by varying the pump laser amplitude.  
As in standard dispersive readout, depending on the frequency shift $\varepsilon = \pm \chi$, $\optfluc_\mathrm{out}(t)$ evolves into one of two different coherent states separated by an angle $2 \xi = 2 \arctan [2 \chi/\Gmech (1 + \Com)]$.
Equation~(\ref{eqn:SNR}) maps to a standard cQED dispersive readout where the cavity damping rate has been replaced by 
an optically-tunable mechanical damping rate $\Gmech (1 + C_\mathrm{om})$, and where only a fraction $C_\mathrm{om}/(1 + C_\mathrm{om})$ of the total output flux is detected. 
As we show, this additional tunability leads to important differences in readout optimization and dynamics.

\prlsection{Measurement time}

\begin{figure}
	\centering
	\includegraphics[width=0.48\textwidth]{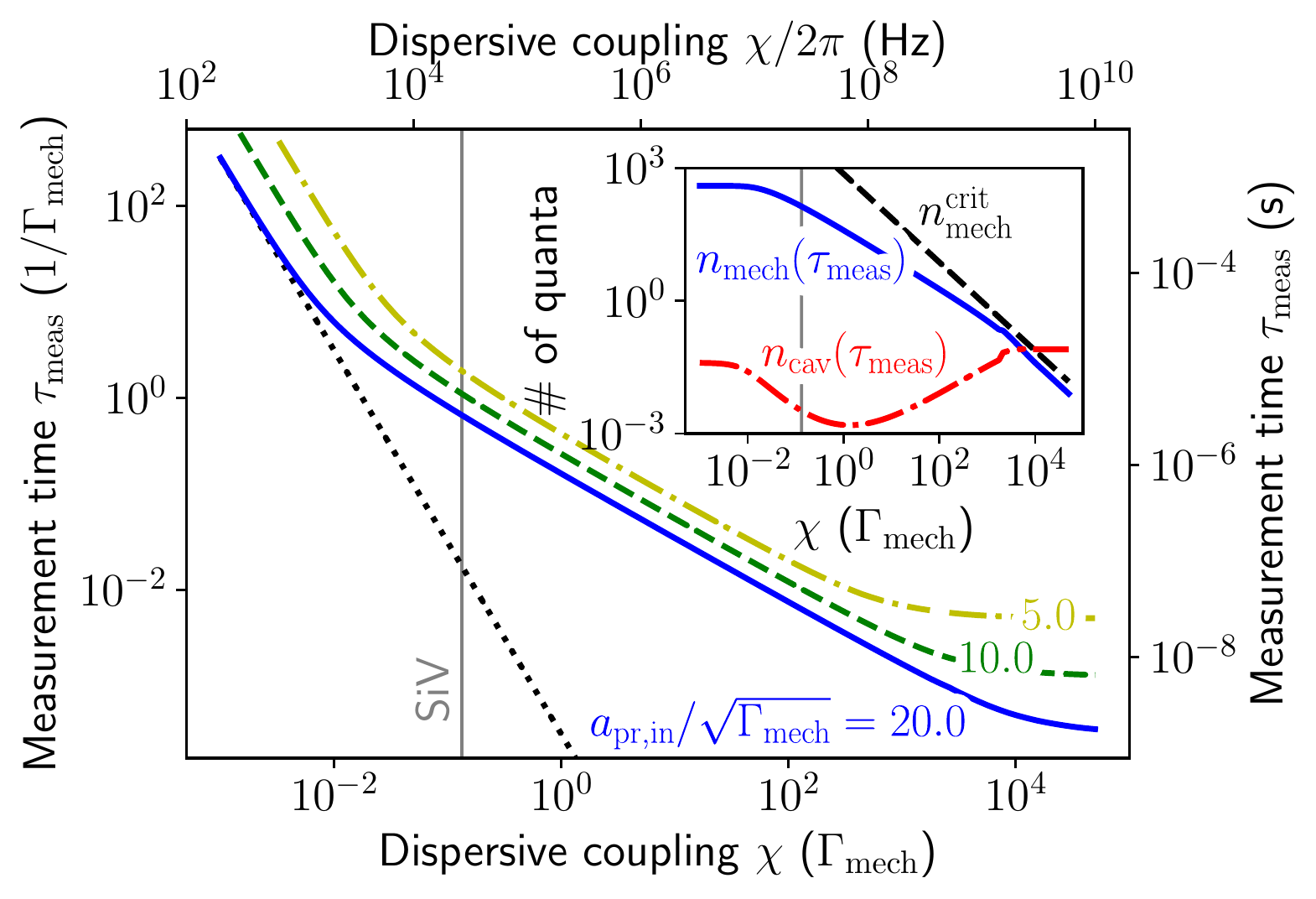}
	\caption{
		Minimum measurement time $\tmeas$ required to reach a signal-to-noise ratio (SNR) of unity as a function of the spin-mechanical dispersive coupling strength $\chi$.
		The optomechanical cooperativity $\Com$ has been optimized for each data point. 
		The dotted black line indicates the asymptotic form of the measurement time for $\chi/\Gmech \ll 1$ [Eq.~\eqref{eqn:tmeasWeakCoupling}].
		It is off by almost two orders of magnitude for the expected parameters for SiV defects in a diamond OMC (gray vertical line). 
		Inset: Phonon number $\nmech(\tmeas)$, photon number $\ncav(\tmeas)$, and critical phonon number $\ncrit$ for $a_\mathrm{pr,in}/\sqrt{\Gmech} = \Valueapr$, $\gspin$ varied, $\Deltaspin/\Gmech = \ValueDeltasmOverGamma$, and $\kappa/\Gmech = \ValueRatiokappaoverGamma$. 
	}
	\label{fig:zerotemperature}
\end{figure}

The measurement time is implicitly defined by $\SNR^2(\tmeas) = 1$, and our goal is to optimize $\Com$ such that $\tmeas$ is minimal.
As shown in Fig.~\ref{fig:zerotemperature}, there are three scalings of $\tmeas$ with $\chi$:
(i) For a weak strain coupling $\chi \ll \Gmech,\kappa$, the intrinsic mechanical ringup time $1/\Gmech$ is much shorter than $\tmeas$. 
The measurement is fastest if the impedance-matching condition $\Com = 1$ holds, in which case
\begin{align}
	\tmeas \to \frac{\Gmech^2}{8 \abs{a_\mathrm{pr,in}}^2 \chi^2} \fullstop
	\label{eqn:tmeasWeakCoupling}
\end{align}
In this regime, the probe laser leads to a  steady-state mechanical phonon number $\nmech^\mathrm{ss} = \lim_{\tau \to \infty} \nmech(\tau) = \abs{a_\mathrm{pr,in}}^2/\Gmech$ on a timescale shorter than $\tmeas$.   
(ii) As we show below, spin defects can reach appreciable strain coupling $\chi \gtrsim \Gmech$ such that $\SNR^2(\tau)=1$ is achieved before the mechanical steady state is reached.
In this regime, is it advantageous to increase $\Com$ beyond $1$ to speed up the mechanical ring-up, so that this occurs on the same timescale as the measurement (i.e., $\Com \propto 1/\Gmech \tmeas$).  
For an optimal $\Com$, we find in this regime $\tmeas \propto (\chi \abs{a_\mathrm{pr,in}})^{-2/3}$. 
(iii) Finally, for $\chi \gg \Gmech$, the large detuning $\pm \chi$ between the mechanical mode and the probe laser becomes the limiting factor of the measurement.
The optimal cooperativity $\Com = 2 \chi/\Gmech$ strongly broadens the mechanical linewidth such that transient dynamics becomes irrelevant again and the measurement time converges to a constant value which depends only on the rate at which probe photons are sent into the system, 
\begin{align}
	\tmeas \to \frac{1}{8 \abs{a_\mathrm{pr,in}}^2} \fullstop
	\label{eqn:FundamentalLimittmeas}
\end{align}
Note that OMIT allows one to optimize the effective damping rate for different values of the dispersive coupling such that one can take advantage of large couplings $\chi \gg \Gmech$.

\prlsection{Critical phonon number}

Figure~\ref{fig:zerotemperature} reveals several interesting features.
First, $\tmeas$ can be smaller than $1/\Gmech$, which reflects the fact that we can broaden the mechanical linewidth optically, $\Gmech (1 + \Com) \gg \Gmech$. 
Second, $\tmeas$ is short because we are using many probe phonons.
As shown in the inset of Fig.~\ref{fig:zerotemperature}, this does not come at the cost of a high photon number (which could cause unwanted heating) since $\nmech^\mathrm{ss}/\ncav^\mathrm{ss} = \Gmech \kappa \Com/(\Gmech^2 + 4 \varepsilon^2) \propto \kappa/\Gmech \gg 1$.
Further, with increasing $\chi/\Gmech$, the optimized $\Com$ grows and  $\nmech^\mathrm{ss}$ decreases (as the total mechanical damping is $\propto \Com$).
Corrections to the dispersive spin-mechanical interaction define a critical phonon number $\ncrit$ (see \cite{SM}), which limits the maximum probe power, determines the plateau value of $\tmeas$ for $\chi \gg \Gmech$, and prevents infinitely fast measurements \cite{FN4}.

\prlsection{Feasibility criteria}

For QND readout, one needs $\tmeas \ll \min(T_1, \tau_\mathrm{Purcell})$, where $T_1 = 2 \pi/\gamma_\mathrm{rel}$ is the single-spin relaxation time and $\tau_\mathrm{Purcell}$ the Purcell decay time. 
As we show below, this is well within reach for a single SiV defect coupled to a diamond OMC. 
For other defects with smaller strain coupling, this condition can still be achieved in an ensemble of $N$ spins. 
In the regime $\chi \ll \Gmech$, one then obtains the conditions $\Deltaspin/\Gmech \gg \sqrt{N/8}$ (to suppress collective Purcell decay) and $4 N \gspin^2/\Gmech \gamma_\mathrm{rel} \gg 1/2$ \cite{SM}.

\prlsection{Application to SiV systems}

As a concrete example, we show that readout of a single SiV defect embedded in a state-of-the-art diamond OMC is experimentally feasible. 
Diamond OMCs with $\kappa/2\pi \approx \Valuekappa$ have recently been demonstrated by Burek \emph{et al.} \cite{Burek2016} and Cady \emph{et al.} \cite{Cady2019}. 
The mechanical modes had $\ommech/2\pi \approx \Valueommech$ and quality factors up to $4100$ at room temperature with higher values expected at cryogenic temperatures \cite{Burek2016}.
A mechanical damping rate $\Gmech/2\pi = \ValueGammaMech$ seems thus feasible. 
The measured optomechanical couplings are $\gom/2\pi \approx \Valuegom$ \cite{Burek2016,Cady2019}. 
Spin-mechanical single-phonon coupling rates for SiV defects in an OMC have been estimated to be $\gspin/2\pi \approx \Valuegsm$ \cite{Shandilya2021}.
Surprisingly, the strain coupling can be tuned up to $\gspin/2\pi \approx 8\,\mathrm{MHz}$ by applying a suitable off-axis magnetic field without changing the SiV level splitting, as we show in a detailed microscopic analysis in the supplemental material \cite{SM}.
Using $\gspin/2\pi = \Valuegsm$ as a conservative estimate and assuming a detuning $\Deltaspin \equiv \ommech - \omspin = 2\pi \times \ValueDeltasm$, a dispersive coupling $\chi \equiv \gspin^2/\Deltaspin= 2\pi \times \Valuechi$ appears to be realistic. 
The corresponding ratio $\chi/\Gmech = \ValuechioverGammaMech$ is indicated by the gray vertical line in Fig.~\ref{fig:zerotemperature}.

With these numbers, and using low probe-laser power [such that $\nmech(\tmeas$) is more than an order of magnitude below $\ncrit$,  see inset of Fig.~\ref{fig:zerotemperature}], we find an estimated measurement time of $\tmeas = \ValueTauMeas$.  
This could be further decreased by using a stronger probe laser.  
Our $\tmeas$ is thus competitive with optical readout times of $13\,\upmu\mathrm{s}$ for highly strained SiV centers in a diamond nanocavity \cite{Nguyen2019} and $30\,\upmu\mathrm{s}$ for optical fluorescence readout of SiV centers with an external magnetic field precisely aligned along the SiV axis. 
In the latter case, the measurement times were limited by the repolarization of the SiV into its ground state on a timescale $\approx 30\,\mathrm{ms}$. 
For OMIT readout, the estimated measurement times are an order of magnitude shorter, and they will be limited by a Purcell decay time of $\tau_\mathrm{Purcell} \approx \ValueTauPurcell$ \cite{SM}.
We thus find $\tau_\mathrm{Purcell}/\tmeas \approx \ValueRatioTauPurcellOverTauMeas \ggg 1$, which could be further increased by increasing $\Deltaspin$ \cite{FN5}.

\prlsection{Application for quantum sensing}

\begin{figure}
	\centering
	\includegraphics[width=0.48\textwidth]{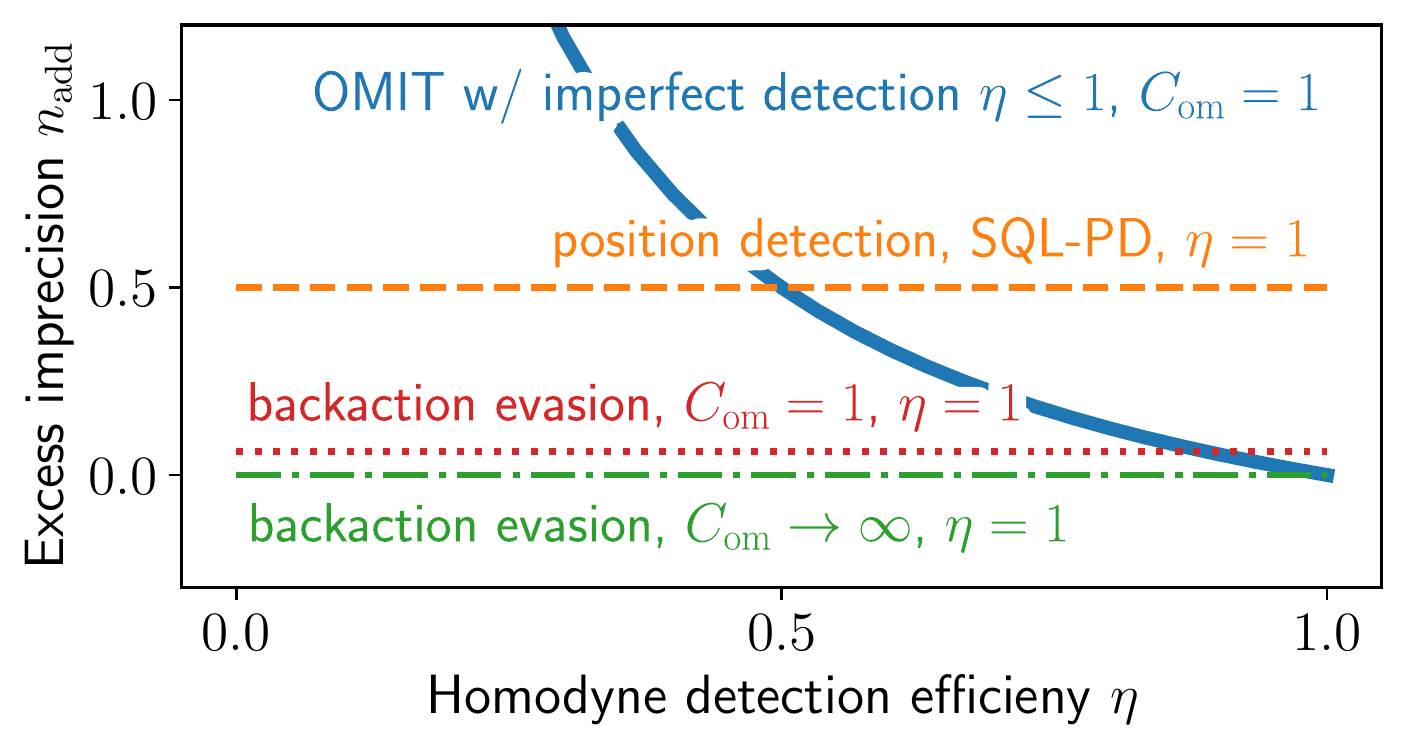}
	\caption{
		    Comparison of the excess imprecision noise $n_\mathrm{add}$ [expressed in terms of an equivalent amount of thermal phonons, see Eq.~\eqref{eqn:Sensitivity}], for sensing a small mechanical frequency shift $\varepsilon \ll \ommech$ with different measurement schemes.
			The thick blue line indicates an OMIT measurement with imperfect homodyne detection (efficiency $0 \leq \eta \leq 1$). 
			The thin horizontal lines indicate the smallest excess imprecision noise achievable with other schemes in the limit of perfect homodyne detection, $\eta = 1$. 
	}
	\label{fig:sensing}
\end{figure}

Our OMIT measurement scheme can also be used for more general parameter estimation where the goal is to detect an unknown signal that causes a small mechanical frequency shift $\varepsilon \ll \ommech$. 
This basic sensing scheme is widely used, e.g., in atomic-force microscopy (AFM) \cite{Martin1987,Albrecht1991} and mass sensing \cite{Ekinci2004}, and it has also been suggested for new OM sensing protocols using limit cycles \cite{Guha2020}.
Here, with quantum sensing in mind, we are interested in the fundamental limits on the estimation error of such schemes. 
OMIT allows one to improve the estimation error beyond that of standard schemes using continuous mechanical position detection. 
Such schemes are fundamentally limited by the standard quantum limit of position detection (SQL-PD) \cite{Braginsky1992,Clerk2010}. 
The estimation error for infinitesimal frequency changes is $(\mathbf{\Delta}\varepsilon)^2(\tau) = \lim_{\varepsilon \to 0} \cerw{[\widehat{\delta \mathcal{I}}(\tau)]^2}_\varepsilon/\cabs{\partial_\varepsilon \cerw{\hat{\mathcal{I}}(\tau)}_\varepsilon}^2$, which is optimized for a resonant probe laser and $\Com = 1$ \cite{SM}, 
\begin{align}
	(\mathbf{\Delta}\varepsilon)^2(\tau) = \frac{\Gmech}{4 \nmech^\mathrm{ss} \tau} (1 + 2 \nth + 2 \nmeas) \fullstop
	\label{eqn:Sensitivity}
\end{align}
Here, $\nth$ denotes the thermal phonon number due to interaction of the mechanical mode with a finite-temperature environment, and $\nmeas$ represents potential imprecision noise due to the readout of the mechanics, expressed in terms of an equivalent amount of thermal phonons.
Note that AFM and limit-cycle sensing protocols yield the same estimation error~\eqref{eqn:Sensitivity} but are limited by thermal noise, $\nth \gg 1$, and thus typically not sensitive to fundamental imprecision noise \cite{Albrecht1991,Guha2020}. 
Also note that our goal is not to change the fundamental scaling with $\nmech^\mathrm{ss}$, but to make $\nmeas$ as small as possible.

In the ideal case analyzed so far, we have $\nmeas = 0$ for OMIT readout even when all quantum effects are included. 
To highlight the significance of this result, it is instructive to compare Eq.~\eqref{eqn:Sensitivity} with other measurement schemes to determine a small frequency shift $\varepsilon$. 
Perhaps the most obvious approach is to drive the mechanical resonator linearly at $\ommech$ and continuously measure its position $\hat{x} = x_\mathrm{zpf} (\mechmode + \mechmode^\dagger)$, where $x_\mathrm{zpf}$ denotes the zero-point fluctuations.
This signal can then be used to determine the phase lag between $\cerw{\hat{x}(t)}$ and the drive (and hence $\varepsilon$). 
Since this measurement collects information on both quadratures of $\hat{x}(t)$, its estimation error can at best reach the SQL-PD with $\nmeas = 1/2$ \cite{Braginsky1992,Clerk2010}.

The SQL-PD can be surpassed by performing a backaction-evading (BAE) measurement \cite{Braginsky1980,Clerk2008}, which is tuned to measure only the phase quadrature of $\hat{x}(t)$ containing information on $\varepsilon$. 
In the limit of a large cooperativity $\Com \to \infty$, one finds $\nmeas \to 0$ and thus achieves \emph{the same estimation error} as our OMIT scheme. 
While BAE measurements (which necessarily require large $\Com$) have been demonstrated \cite{Hertzberg2010,Suh2014,Lecocq2015,OckeloenKorppi2016,Shomroni2019}, they are experimentally far more challenging than a simple OMIT measurement with $\Com = 1$ (something that is routinely done for characterization purposes). 

Note that both direct position detection and BAE measurements require careful phase tuning between the mechanical drive and the local oscillator of the homodyne detection.
In contrast, our OMIT scheme is an all-optical measurement where the optical probe (driving the mechanics) and the local oscillator can be derived from the same laser, eliminating the need for a separate mechanical drive and its phase control.

The absence of added noise in the OMIT scheme is due to the fact that OMIT (unlike position detection and BAE measurements) transduces both mechanical quadratures into quadratures of the optical output field \emph{without any gain} \cite{SM}. 
By adjusting the local-oscillator phase, one can then choose to measure the optical quadrature $\propto \varepsilon$. 
Amplification of mechanical quadratures is not required since once can increase the signal by driving the mechanics more strongly, which gives rise to the $1/\nmech^\mathrm{ss}$ scaling in Eq.~\eqref{eqn:Sensitivity}.
In Fig.~\ref{fig:sensing}, we also analyze the case of imperfect homodyne detection (efficiency $0 \leq \eta \leq 1$).
In this case, there will be added noise $\nmeas = (1 - \eta)/2 \eta$, but state-of-the-art OMIT detection will surpass the SQL-PD for experimentally feasible efficiencies $\eta \gtrsim 70\,\%$ \cite{Purdy2013}.

\prlsection{Conclusion}

Our work presents a potentially powerful  alternative readout scheme for solid-state spin defects with large strain coupling.
This coupling allows one to perform dispersive spin readout using a mechanical mode, which is optically driven and read out using an OMIT scheme \cite{FN6}.
For SiV defects in a diamond OMCs, the estimated readout times are an order of magnitude shorter than the best measurement times for single-shot optical fluorescence readout. 
Besides spin readout, our scheme is also useful for quantum sensing, when a small signal modifies the resonance frequency of a mechanical oscillator, e.g., strain-mediated readout of the collective state of a large ensemble of NV centers.
It would be interesting to check if OMIT readout can also be applied to other types of solid-state spin defects with strain coupling.

Our protocol could be combined with existing ideas to generate remote entanglement between two distant superconducting qubits using dispersive measurements \cite{Roch2014,Silveri2016}, requiring only small modifications of recent experiments coupling superconducting qubits to mechanical modes \cite{vonLuepke2022,Wollack2022,Mirhosseini2020,FN7}.

\let\oldaddcontentsline\addcontentsline
\renewcommand{\addcontentsline}[3]{}

\begin{acknowledgments}
This work was supported by the Defense Advanced Research Projects Agency (DARPA) Driven and Nonequilibrium Quantum Systems (DRINQS) program (Agreement D18AC00014). 
We also acknowledge support from the DOE Q-NEXT Center (Grant No. DOE 1F-60579), the NSF QLCI program (Grant No. OMA-2016245), and from the Simons Foundation (Grant No. 669487, A.~C.). 
C.~P.\ acknowledges support from the NSF Quantum Foundry at UCSB (NSF DMR-1906325).
J.~V.~C.\ acknowledges support from NASA STTR (Contract 80NSSC22PB210).
\end{acknowledgments}

\bibliography{citations}

\begin{thebibliography}{90}
\bibitem [{FN1()}]{FN1}%
  \BibitemOpen
  \href@noop {} {}\bibinfo {note} {While there are many differences \cite{SM},
  the most crucial is that our scheme has fewer basic ingredients: there is no
  microwave resonator or drive, nor any microwave-optomechanical
  coupling.}\BibitemShut {Stop}%
\bibitem [{FN4()}]{FN4}%
  \BibitemOpen
  \href@noop {} {}\bibinfo {note} {Here, we always choose the optical probe
  power small enough to keep $\nmech^\mathrm{ss}$ below the critical phonon
  number. Thanks to the extreme linearity of mechanical modes in OMCs and the
  small bare optomechanical coupling strength, $\nmech^\mathrm{ss} \gg 1$ does
  not lead to a breakdown of the linearized optomechanical
  Hamiltonian.}\BibitemShut {Stop}%
\bibitem [{FN5()}]{FN5}%
  \BibitemOpen
  \href@noop {} {}\bibinfo {note} {Given such a large value of $\chi/\Gmech$,
  one may worry that the spin-mechanical coupling degrades the coherence
  properties of the SiVs. However, resonant interactions with phonon modes
  other than the mechanical mode of interest can be suppressed by a suitable
  design of the OMC. Dephasing of the SiV is irrelevant for spin readout but it
  may still limit the operation of the SiV as a quantum sensor. Yet, the main
  phonon-induced dephasing mechanism is dispersive coupling to off-resonant
  several-GHz phonon modes, which will be highly suppressed at temperatures
  below $100\,\mathrm{mK}$ \cite{SM}.}\BibitemShut {Stop}%
\end{thebibliography}
\let\addcontentsline\oldaddcontentsline

\newpage 
\clearpage
\thispagestyle{empty}
\onecolumngrid
\begin{center}
\textbf{\large Supplemental Material for\\\thetitle}
\end{center}

\begin{center}
\theauthors\\
\emph{\theaffiliations}\\
(Dated: \today)
\end{center}

\setcounter{equation}{0}
\setcounter{figure}{0}
\setcounter{table}{0}
\setcounter{page}{1}
\makeatletter
\renewcommand{\theequation}{S\arabic{equation}}
\renewcommand{\thefigure}{S\arabic{figure}}
\renewcommand{\bibnumfmt}[1]{[S#1]}
\renewcommand{\citenumfont}[1]{#1}


\tableofcontents

\section{Dispersive spin readout using OMIT}
\subsection{Optical and mechanical output fields}
\label{sec:SM:OutputFields}

We consider an optomechanical system with an optical mode $\optmode$ (at frequency $\omopt$) and a mechanical mode $\mechmode$ (at frequency $\ommech$), which are coupled with a bare optomechanical coupling strength $\gom$, 
\begin{align}
	\hat{H}_\mathrm{om} &= \omopt \optmode^\dagger \optmode + \ommech \mechmode^\dagger \mechmode - \gom \optmode^\dagger \optmode \left( \mechmode^\dagger + \mechmode \right) \fullstop
\end{align}
Here and in the following, we set $\hbar = 1$. 
The optical mode is driven by a strong pump laser at frequency $\ompump = \omopt - \ommech$, 
\begin{align}
	\hat{H}_\mathrm{pump} &= \sqrt{\kappa} \left[ a_\mathrm{pump,in}(t) \optmode^\dagger + a_\mathrm{pump,in}^*(t) \optmode \right] \comma
\end{align}
where $a_\mathrm{pump,in}(t) = \bar{a}_\mathrm{pump,in} e^{-i \ompump t}$ is the semiclassical part of the input field due to the pump laser and 
$\kappa$ is the decay rate of the optical mode due to coupling to the input/output channel. 
The photon flux of the pump laser is given by $\abs{\bar{a}_\mathrm{pump,in}}^2$. 
The mechanical mode is coupled to a single spin-$1/2$ system (with level-splitting energy $\omspin$) by a Jaynes-Cummings interaction with spin-mechanical coupling strength $\gspin$, 
\begin{align}
	\hat{H}_\mathrm{JC} &= \frac{\omspin}{2} \hat{\sigma}_z + \gspin \left( \hat{\sigma}_+ \mechmode + \hat{\sigma}_- \mechmode^\dagger \right) \fullstop 
\end{align}
Here, $\hat{\sigma}_{x,y,z}$ are the Pauli matrices, and $\hat{\sigma}_\pm = (\hat{\sigma}_x \pm i \hat{\sigma}_y)/2$ are the spin raising and lowering operators, respectively. 
Both the optical and the mechanical mode are coupled to a dissipative environment modeled by a Lindblad quantum master equation,
\begin{align}
	\frac{\d}{\d t} \hat{\rho} 
		&= -i \komm{\hat{H}_\mathrm{om} + \hat{H}_\mathrm{JC} + \hat{H}_\mathrm{pump}}{\hat{\rho}} 
		+ \kappa \mathcal{D}[\optmode] \hat{\rho} 
		+ \Gmech (\nth + 1) \mathcal{D}[\mechmode] \hat{\rho} + \Gmech \nth \mathcal{D}[\mechmode^\dagger] \hat{\rho} \comma
\end{align}
where $\Gmech$ is the bare mechanical decay rate, $\nth$ is the thermal phonon number, and the Lindblad dissipators are defined as $\mathcal{D}[\hat{O}]\hat{\rho} = \hat{O} \hat{\rho} \hat{O}^\dagger - \cakomm{\hat{O}^\dagger \hat{O}}{\hat{\rho}}/2$. 

To simplify the optomechanical Hamiltonian, we switch to a rotating frame at the pump laser frequency $\ompump$ for the optical mode, decompose the optical (mechanical) field into a large semiclassical amplitude $a$ ($b$) and quantum fluctuations $\optfluc$ ($\mechfluc$) around it, 
\begin{align}
	\optmode &= e^{-i \ompump t} \left( a + \optfluc \right) \comma \\
	\mechmode &= b + \mechfluc \comma
\end{align}
and linearize the optomechanical Hamiltonian \cite{Aspelmeyer2014RMP}. 
The semiclassical amplitudes $a$ and $b$ are the self-consistent solutions of 
\begin{align}
	a &= \frac{\sqrt{\kappa} \bar{a}_\mathrm{pump,in}}{\Deltaopt + g_0 (b + b^*) + i \frac{\kappa}{2}} \comma &
	b &= \frac{g_0 \abs{a}^2}{\ommech - i \frac{\Gmech}{2}} \comma
\end{align}
where we introduced the detuning $\Deltaopt = \ompump - \omopt$.
Assuming $a$ to be real without loss of generality and assuming the pump laser to be red-detuned with respect to the cavity resonance frequency, $\Deltaopt \approx - \ommech$, we obtain the linearized equation of motion
\begin{align}
	\frac{\d}{\d t} \hat{\rho}
		&= - i \komm{\hat{H}_\mathrm{om}' + \hat{H}_\mathrm{JC}'}{\hat{\rho}} + \kappa \mathcal{D}[\optfluc]\hat{\rho} + \Gmech (\nth + 1) \mathcal{D}[\mechfluc] \hat{\rho} + \Gmech \nth \mathcal{D}[\mechfluc^\dagger] \hat{\rho} \comma \\
	\hat{H}_\mathrm{om}' &= - \Deltaoptbar \optfluc^\dagger \optfluc + \ommech \mechfluc^\dagger \mechfluc - G \left( \optfluc^\dagger \mechfluc + \mechfluc^\dagger \optfluc \right) \comma 
	\label{eqn:SM:system:LinearizedHom} \\
	\hat{H}_\mathrm{JC}' &= \frac{\omspin}{2} \hat{\sigma}_z + \gspin \left[ (b + \mechfluc) \hat{\sigma}_+ + (b^* + \mechfluc^\dagger) \hat{\sigma}_- \right] \comma
\end{align}
where we introduced the renormalized detuning $\Deltaoptbar = \Deltaopt + g_0 (b + b^*)$ and the optically-enhanced coupling strength $G = g_0 a$. 
Note that the renormalization of the optical frequency by the mechanical steady-state amplitude, $\Deltaopt \to \Deltaoptbar = \Deltaopt + g_0 (b + b^*)$, is very small in the case considered here: 
For the experimental parameters and values of $\Com$ considered in the main text, we find $\abs{g_0 (b + b^*)}/\abs{\Deltaopt} \approx 10^{-6} \Com \ll 1$.

We assume that the spins are strongly detuned from the mechanical mode, $\abs{\Deltaspin} \equiv \abs{\ommech - \omspin} \gg \gspin$, such that the spin-mechanical interaction can be diagonalized up to second order in $\gspin/\Deltaspin$ by a Schrieffer-Wolff transformation $\hat{H}'= e^{\hat{S}} \hat{H} e^{- \hat{S}}$ with 
\begin{align}
	\hat{S} = \frac{\gspin}{\Deltaspin} \left[ \left( \mechfluc^\dagger - \frac{\Deltaspin}{\omspin} b \right) \hat{\sigma}_- - \left( \mechfluc - \frac{\Deltaspin}{\omspin} b^* \right) \hat{\sigma}_+ \right] \fullstop
	\label{eqn:SM:system:SchriefferWolff}
\end{align}
To leading order in $\gspin/\Delta$, ignoring constant terms, and using a rotating wave approximation, we find
\begin{align}
	\frac{\d}{\d t} \hat{\rho}
		&= -i \komm{\hat{H}_\mathrm{om}' + \hat{H}_\mathrm{sm}'}{\hat{\rho}} + \kappa \mathcal{D}[\optfluc] \hat{\rho} 
		+ \Gmech (\nth + 1) \mathcal{D}[\mechfluc] \hat{\rho} + \Gmech \nth \mathcal{D}[\mechfluc^\dagger] \hat{\rho} \nonumber \\
		&\phantom{=}\ + \frac{\Gmech \chi}{\Deltaspin} (\nth + 1) \mathcal{D}[\hat{\sigma}_-] \hat{\rho} + \frac{\Gmech \chi}{\Deltaspin} \nth \mathcal{D}[\hat{\sigma}_+] \hat{\rho} \comma 
		\label{eqn:SM:system:QME} \displaybreak[1]\\
	\hat{H}_\mathrm{sm}' 
		&= \left( \frac{\omspin - \chi}{2} + \frac{\chi \Deltaspin}{\omspin} \abs{b}^2 \right) \hat{\sigma}_z - \chi \hat{\sigma}_z \mechfluc^\dagger \mechfluc \comma
\end{align}
where we defined the dispersive coupling strength $\chi = \gspin^2/\Deltaspin$.
For typical experimental parameters, the conditions $\chi \ll \omspin$ and $\chi \Deltaspin \abs{b}^2/\omspin^2 \ll 1$ hold and we can neglect the correction terms to the spin-transition frequency, $\hat{H}_\mathrm{sm}' \approx \hat{H}_\mathrm{sm} = \omspin \hat{\sigma}_z/2 - \chi \hat{\sigma}_z \mechfluc^\dagger \mechfluc$ (which is the form of the spin-mechanical Hamiltonian given in the main text).

Note that, apart from the Purcell decay terms in Eq.~\eqref{eqn:SM:system:QME}, $\hat{\sigma}_z$ is a constant of motion. 
The Purcell decay time $\tau_\mathrm{Purcell} \propto \Deltaspin/\Gmech \chi$ can be made arbitrarily large by increasing the spin-mechanical detuning $\Deltaspin$, such that we implement a quantum-nondemolition (QND) measurement.
We can therefore ignore the spin dynamics and only include the constant spin-state-dependent mechanical frequency shift $\varepsilon = - \chi \cerw{\hat{\sigma}_z}$ in the optomechanical Hamiltonian, 
\begin{align}
	\hat{H}_\mathrm{om}' \to - \Deltaoptbar \optfluc^\dagger \optfluc + (\ommech + \varepsilon) \mechfluc^\dagger \mechfluc - G \left( \optfluc^\dagger \mechfluc + \mechfluc^\dagger \optfluc \right) 
	\label{eqn:SM:system:LinearizedHomWithSpinShift}
\end{align}
The linearized optomechanical Hamiltonian given in Eq.~\eqref{eqn:LinearizedHamiltonian} of the main text is Eq.~\eqref{eqn:SM:system:LinearizedHomWithSpinShift} for $\Deltaoptbar = - \ommech$.

From Eqs.~\eqref{eqn:SM:system:QME} and~\eqref{eqn:SM:system:LinearizedHomWithSpinShift}, we obtain the following Heisenberg-Langevin equations of motion for the mechanical fluctuations $\mechfluc$ and the optical fluctuations $\optfluc$. 
\begin{align}
	\frac{\d}{\d t} \optfluc &= - \left[ \frac{\kappa}{2} - i \Deltaoptbar \right] \optfluc + i G \mechfluc - \sqrt{\kappa} \optfluc_\mathrm{in} \comma 
	\label{eqn:SM:system:Langevin:d}\\
	\frac{\d}{\d t} \mechfluc &= - \left[ \frac{\Gmech}{2} + i (\ommech + \varepsilon) \right] \mechfluc + i G \optfluc - \sqrt{\Gmech} \mechfluc_\mathrm{in} \comma
	\label{eqn:SM:system:Langevin:b}
\end{align}
where $\optfluc_\mathrm{in}$ and $\mechmode_\mathrm{in}$ are the optical and mechanical input fields, respectively.

The optical input field on the right-hand-side of Eq.~\eqref{eqn:SM:system:Langevin:d} contains the semiclassical field of the weak probe laser (which is switched on instantaneously at $t=0$) as well as zero-temperature Gaussian white noise $\hat{\xi}_\mathrm{in}$, 
\begin{align}
	\optfluc_\mathrm{in}(t) &= a_\mathrm{pr,in} e^{-i \omprrot t} \Theta(t) + \hat{\xi}_\mathrm{in}(t) \comma 
	\label{eqn:SM:system:din}\\
	\erw{\hat{\xi}_\mathrm{in}(t) \hat{\xi}_\mathrm{in}^\dagger(t')} &= \delta(t - t') \comma 
\end{align} 
where $\Theta(t)$ denotes the Heaviside step function and $\abs{a_\mathrm{pr,in}}^2$ is the photon flux due to the probe laser. 
Note that $\optfluc$ and $\optfluc_\mathrm{in}$ are defined in a frame rotating at $\ompump$, i.e., the probe frequency in the lab frame is given by $\ompr = \ompump + \omprrot = \omopt - \ommech + \omprrot$ and the probe laser is on resonance with the optical cavity if $\omprrot = \ommech$. 
The mechanical input field in Eq.~\eqref{eqn:SM:system:Langevin:b} is given by finite-temperature Gaussian white noise,
\begin{align}
	\erw{\mechfluc_\mathrm{in}(t) \mechfluc_\mathrm{in}^\dagger(t')} &= (\nth + 1) \delta(t - t') \comma 
	\label{eqn:SM:system:mechanicalnoiseCorr1} \\
	\erw{\mechfluc_\mathrm{in}^\dagger(t) \mechfluc_\mathrm{in}(t')} &= \nth \delta(t - t') \fullstop	
	\label{eqn:SM:system:mechanicalnoiseCorr2}
\end{align}
The Heisenberg-Langevin equations~\eqref{eqn:SM:system:Langevin:d} and~\eqref{eqn:SM:system:Langevin:b} can be solved exactly and, for $\Deltaoptbar = - \ommech$, their solution is
\begin{align}
	\begin{pmatrix}
		\optfluc(t) \\ \
		\mechfluc(t)
	\end{pmatrix} = - \int_{-\infty}^\infty \d t'\ \mathcal{G}(t - t') \cdot \begin{pmatrix}
		\sqrt{\kappa} & 0 \\
		0 & \sqrt{\Gmech}
	\end{pmatrix} \cdot \begin{pmatrix}
		\optfluc_\mathrm{in}(t') \\
		\mechfluc_\mathrm{in}(t') 
	\end{pmatrix} \comma
	\label{eqn:SM:system:exactSolutionFordAndb}
\end{align}
where $\mathcal{G}(\tau)$ denotes the Green's function of the Heisenberg-Langevin equations, 
\begin{align}
	\mathcal{G}(\tau) &= \Theta(\tau) e^{-i \tau (\ommech + \varepsilon/2) - (\Gmech + \kappa)/4} \begin{pmatrix}
		G_{\optflucsymbol \optflucsymbol}(\tau) & G_{\optflucsymbol \mechflucsymbol}(\tau) \\
		G_{\mechflucsymbol \optflucsymbol}(\tau) & G_{\mechflucsymbol \mechflucsymbol}(\tau)
	\end{pmatrix} \comma \nonumber \\
	G_{\optflucsymbol \optflucsymbol}(\tau) &= \cosh \left[ \frac{\tau \sqrt{\dots}}{4} \right] + \frac{\Gmech - \kappa + 2 i \varepsilon}{\sqrt{\dots}} \sinh \left[ \frac{\tau \sqrt{\dots}}{4} \right] \comma \nonumber\\
	G_{\mechflucsymbol \mechflucsymbol}(\tau) &= \cosh \left[ \frac{\tau \sqrt{\dots}}{4} \right] - \frac{\Gmech - \kappa + 2 i \varepsilon}{\sqrt{\dots}} \sinh \left[ \frac{\tau \sqrt{\dots}}{4} \right] \comma \nonumber\\
	G_{\optflucsymbol \mechflucsymbol}(\tau) &= G_{\mechflucsymbol \optflucsymbol}(\tau) = \frac{2 i \sqrt{\Com \Gmech \kappa}}{\sqrt{\dots}} \sinh \left[ \frac{\tau \sqrt{\dots}}{4} \right] \fullstop 
\end{align}
Here, we introduced the optomechanical cooperativity $\Com = 4 G^2/\kappa \Gmech$ and the abbreviation $\sqrt{\dots} = \sqrt{\Gmech^2 - (2 \varepsilon + i \kappa)^2 - 2 \Gmech [- 2 i \varepsilon + \kappa (1 + 2 C_\mathrm{om})]}$. 
From this result, we can calculate the optical and mechanical output fields using standard input-output theory
\begin{align}
	\optfluc_\mathrm{out}(t) &= \sqrt{\kappa} \optfluc(t) + \optfluc_\mathrm{in}(t) \comma \\
	\mechfluc_\mathrm{out}(t) &= \sqrt{\Gmech} \mechfluc(t) + \mechfluc_\mathrm{in}(t) \fullstop
	\label{eqn:SM:Solution:doutbout}
\end{align}

\subsection{Signal-to-noise ratio}
\label{sec:SM:SNR}

The information on the spin's $\sigma_z$ projection is encoded in the phase shift of the optical output field at the frequency of the probe laser. 
It can be read out by homodyne detection of a suitable quadrature of the optical output field, and integrating the output signal for a time $\tau$. 
Such a measurement is described by the observable
\begin{align}
	\hat{\I}(\tau) = \sqrt{\kappa} \int_0^\tau \d t\, \left[ e^{i \varphi} e^{-i \omprrot t} \optfluc_\mathrm{out}^\dagger(t) + e^{-i \varphi} e^{i \omprrot t} \optfluc_\mathrm{out}(t) \right] \comma
	\label{eqn:SM:IntegratedHomodyneCurrentOperator}
\end{align}
where the angle $\varphi$ determines the measured quadrature and will be specified later.
For the two eigenstates $\hat{\sigma}_z \ket{\sigma_z} = \sigma_z \ket{\sigma_z}$ with $\sigma_z \in \{+1,-1\}$, the optical output field will evolve into two different coherent states.
They give rise to different integrated homodyne currents $\cerw{\hat{\I}(\tau)}_\varepsilon$, where $\cerw{\cdot}_\varepsilon$ denotes an expectation value with the mechanical resonance frequency being shifted by $\varepsilon = - \chi \sigma_z$. 
Since the instantaneous homodyne current fluctuates about its expectation value,  
this signal will be accompanied by state-dependent noise $\cerw{[\widehat{\delta \I}(\tau)]^2}_\varepsilon$, where $\widehat{\delta\I}(\tau) = \hat{\I\\
}(\tau) - \cerw{\hat{\mathcal{I}}(\tau)}_\varepsilon$. 
The signal-to-noise ratio (SNR) of the spin readout process is then given by
\begin{align}
	\SNR^2(\tau) 
	= \frac{\mathrm{S}^2(\tau)}{\mathrm{N}^2(\tau)}	
	= \frac{\abs{ \cerw{\hat{\I}(\tau)}_{-\chi} - \cerw{\hat{\I}(\tau)}_{+\chi}}^2}{\cerw{[\widehat{\delta \I}(\tau)]^2}_{-\chi} + \cerw{[\widehat{\delta \I}(\tau)]^2}_{+\chi}} \comma
\end{align}
which can be evaluated using Eqs.~\eqref{eqn:SM:Solution:doutbout} and~\eqref{eqn:SM:IntegratedHomodyneCurrentOperator}. 
The exact expressions for the signal-to-noise ratio are quite lengthy but they can be simplified by the following observation. 
After switching on the weak probe laser~\eqref{eqn:SM:system:din}, the field in the optical cavity will build up on a timescale $1/\kappa \ll 1/\Gmech$. 
During this time, the mechanical mode is still at rest and the optical output signal carries no information on the mechanical frequency shift $\varepsilon$. 
Information on $\varepsilon$ will only start to be present in the optical output signal when the mechanical motion rings up. 
We can get rid of the short-time dynamics of the cavity field by taking the usual limit $\kappa \gg G, \chi, \Gmech$ while keeping the optomechanical cooperativity $\Com$ fixed. 
This simplifies the expressions significantly and we find
\begin{align}
	\abs{\cerw{\hat{\I}(\tau)}_{-\chi} - \cerw{\hat{\I}(\tau)}_{+\chi}}^2
		&= \frac{8 \sqrt{\kappa} \abs{a_\mathrm{pr,in}} \Gmech \Com}{\abs{z_-}^2 \abs{z_+}^2} \abs{z_-^2 - (z_+^*)^2 - 2 i \chi \tau z_- z_+^* + (z_+^*)^2 e^{- z_- \tau/2} - z_-^2 e^{- (z_+^*) \tau/2}} \nonumber \\
		&\phantom{=}\ \times \cos \left[ \zeta(\tau) - \varphi + \arg( a_\mathrm{pr,in}) \right] \comma \\
	\cerw{[\widehat{\delta \I}(\tau)]^2}_{\pm \chi} 
		&= \kappa \left[ \tau - \frac{4 \Gmech \Com \nth}{1 + \Com} \frac{(z_\mp^*)^2 \left( 2 - 2 e^{- z_\mp \tau/2} - z_\mp \tau \right) + z_\mp^2 \left( 2 - 2 e^{- z_\mp^* \tau/2} - z_\mp^* \tau \right)}{\abs{z_\mp}^4} \right] \comma \\
	\zeta(\tau) 
		&= \arg \left[ \frac{\sqrt{\kappa} \Gmech \Com \abs{a_\mathrm{pr,in}}}{(z_-)^2 (z_+^*)^2} \left( z_-^2 - (z_+^*)^2 - 2 i \chi \tau z_- z_+^* + (z_+^*)^2 e^{- z_- \tau/2} - z_-^2 e^{- z_+^* \tau/2} \right) \right] \comma
\end{align}
where we used the abbreviation $z_\pm = \Gmech (1 + \Com) + 2 i (\chi \pm \delta)$ and defined the detuning $\delta = \ompr - \omopt$ between the probe laser and the optical cavity resonance frequency. 
Note that the remaining prefactors $\sqrt{\kappa}$ and $\kappa$ are due to the prefactor in Eq.~\eqref{eqn:SM:IntegratedHomodyneCurrentOperator} and will cancel in the signal-to-noise ratio. 
The SNR is thus independent of the optical decay rate $\kappa$, as expected.

The SNR is maximized if one measures the homodyne quadrature $\varphi(\tau) = \zeta(\tau) - \arg(a_\mathrm{pr,in})$ and if the weak probe laser is resonant with the optical cavity, $\delta = 0$.  
In this case, signal and noise take the following simple expressions.
\begin{align}
	\mathrm{S}(\tau) 
		&= 4 \sqrt{\kappa} \abs{a_\mathrm{pr,in}} \frac{\Com}{1 + \Com} \tau \sin(2 \xi) \left[ 1 - F(\tau) \right] \comma 
		\label{eqn:SM:signalStau}\\
	F(\tau) 
		&= \frac{1}{\chi \tau} \left[ \sin(2 \xi) - \sin(2 \xi + \chi \tau) e^{- \Gmech (1 + \Com) \tau/2} \right] \comma 
		\label{eqn:SM:helperFtau}\\
	\mathrm{N}(\tau)
		&= \sqrt{2 \kappa \tau \left[ 1 - G(\tau) \right]} \comma 
		\label{eqn:SM:noiseNtau}\\
	G(\tau)
		&= 8 \nth \Gmech \frac{\Com}{1 + \Com} \frac{2 \cos(2 \xi) - 2 \chi \tau \cos(\xi)/\sin(\xi) - 2 \cos(2 \xi + \chi \tau) e^{- \Gmech (1 + \Com) \tau/2}}{4 \chi^2 \tau /\sin^2(\xi)} \comma 
		\label{eqn:SM:helperGtau}\\
	\xi
		&= \arctan \left[ \frac{2 \chi}{\Gmech (1 + \Com)} \right] \fullstop
		\label{eqn:SM:xi}
\end{align}
Note that Eqs.~\eqref{eqn:SM:signalStau} and~\eqref{eqn:SM:helperFtau} are very similar to the corresponding results for dispersive qubit readout in cavity QED [see, e.g., \cite{Didier2015} for a detailed discussion but note that Eq.~(S15) in the supplemental material of \cite{Didier2015} contains a typo and the correct prefactor of the square brackets should be $1/\chi \tau$]. 
The main difference between our results and the ones in circuit QED are:
\begin{itemize}
	\item The signal~\eqref{eqn:SM:signalStau} has an additional prefactor $\Com/(1+\Com)$, which captures the fact that we can tune how much information on the mechanical mode is contained in the optical output field by changing the amplitude of the pump laser.
	\item The fixed decay rate of the microwave cavity has been replaced by the optically tuneable total mechanical damping rate $\Gmech (1 + \Com)$. 
	\item The mechanical response filters the white mechanical input noise given by Eqs.~\eqref{eqn:SM:system:mechanicalnoiseCorr1} and~\eqref{eqn:SM:system:mechanicalnoiseCorr2}, such that the noise in the optical output field is no longer white if $\nth \neq 0$, see Eq.~\eqref{eqn:SM:helperGtau}. 
\end{itemize} 
To find the shortest measurement rate, we need to maximize the SNR with respect to the optomechanical cooperativity $\Com$ (i.e., the pump laser amplitude).
There will be a nontrivial optimum since the optical output field does not contain any information on $\varepsilon$ both for very small and for very large cooperativities: 
For $\Com \to 0$ mechanical and optical mode are decoupled whereas, for $\Com \to \infty$, the mechanical mode is strongly damped and does not oscillate.

Note that, at short times and for $\nth = 0$, the SNR scales $\propto \tau^{5/2}$ because the mechanics needs to ring up before its state-dependent rotation can discriminate between the two spin states. 
The square brackets in Eq.~\eqref{eqn:SM:IntegratedHomodyneCurrentOperator} thus grow $\propto \tau^2$. 
This result is integrated once with respect to time and divided by the noise, which is purely diffusive at zero temperature, i.e., $\propto \sqrt{\tau}$.

\subsection{Critical phonon number}
\label{sec:SM:MaximumMechanicalPhononNumber}

We now analyze the validity of the approximations made in the derivation of the SNR, given by Eqs.~\eqref{eqn:SM:signalStau} to~\eqref{eqn:SM:xi}. 
From Eq.~\eqref{eqn:SM:system:exactSolutionFordAndb}, we obtain the mechanical phonon number $\nmech$ and the intracavity photon number $\ncav$ due to the optical probe laser, which are given by the following expressions for $\delta = 0$ and $\kappa \gg G, \chi, \Gmech$:
\begin{align}
	\nmech(t) 
		&= \abs{\cerw{\mechfluc(t)}_\varepsilon}^2 
		= \frac{4 \Gmech \Com \abs{a_\mathrm{pr,in}}^2}{\Gmech^2(1 + \Com)^2 + 4 \varepsilon^2} \left( 1 + e^{- \Gmech (1 + \Com) t} - 2 \cos(\varepsilon t) e^{- \Gmech (1 + \Com) t/2} \right) \comma 
		\label{eqn:SM:nmech} \\
	\ncav(t) 
		&= \abs{\cerw{\optfluc(t)}_\varepsilon}^2 
		=  \frac{4 \abs{a_\mathrm{pr,in}}^2}{\kappa \left[ \Gmech^2 (1 + \Com)^2 + 4 \varepsilon^2 \right]} \Big[ \begin{aligned}[t]
				&\Gmech \Com e^{- \Gmech (1 + \Com) t/2} \left( 2 \Gmech \cos(\varepsilon t) - 4 \varepsilon \sin(\varepsilon t) \right) \\
				&+ \Gmech^2 + 4 \varepsilon^2 + \Gmech^2 \Com^2 e^{- \Gmech(1 + \Com) t} \Big] \fullstop
			\end{aligned}
		\label{eqn:SM:nopt}
\end{align}
Note that their ratio is $\nmech(t)/\ncav(t) \propto \kappa/\Gmech \gg 1$, i.e., even a small intracavity photon number leads to a very large mechanical phonon number. 
Various assumptions of our derivation in Sec.~\ref{sec:SM:OutputFields} may break down if the mechanical phonon number becomes too large:

First, nonlinearities of the mechanical mode may become relevant if the root-mean-square (RMS) amplitude of oscillation becomes comparable to the dimensions of the OMC. 
Given zero-point fluctuations $x_\mathrm{zpf} \approx 1 - 10\,\mathrm{fm}$ and devices of the order of micrometers \cite{Burek2016,Aspelmeyer2014}, mechanical nonlinearities will only show up at extremely high phonon numbers $\nmech \approx \mathcal{O}(10^{16})$.

Second, already at lower phonon numbers, the linearization of the optomechanical interaction used to derive Eq.~\eqref{eqn:SM:system:LinearizedHom} may break down. 
In a frame where both $\optfluc$ and $\mechfluc$ oscillate at $\ommech$, the counter-rotating terms neglected in Eq.~\eqref{eqn:SM:system:LinearizedHom} are
\begin{align}
	\hat{H}_\mathrm{om}^\mathrm{cr} = - G \left( e^{-2 i \ommech t} \optfluc \mechfluc + e^{2 i \ommech t} \optfluc^\dagger \mechfluc^\dagger \right) - g_0 \optfluc^\dagger \optfluc \left( e^{-i \ommech t} \mechfluc + e^{i \ommech t} \mechfluc^\dagger \right) \fullstop
	\label{eqn:SM:RotatingTerms}
\end{align}
Their contribution to the dynamics is negligible if the conditions 
\begin{align}
	G \sqrt{\nmech} &\ll 2 \ommech &
	&\text{and} &
	g_0 \sqrt{\nmech} &\ll \ommech
\end{align}
hold, which are equivalent to $\nmech \ll 10^6/\Com$ and $\nmech \ll 10^9$, respectively. 
Both conditions are satisfied for our parameters, as shown by the inset of Fig.~\ref{fig:zerotemperature} in the main text.

Third, we need to ensure that higher-order corrections to the dispersive spin-mechanical coupling remain negligible. 
This turns out to be the most restrictive constraint on our system. 
From Eq.~\eqref{eqn:SM:system:SchriefferWolff}, we see that higher-order terms in the Schrieffer-Wolff transformation remain negligible if the condition 
\begin{align}
	\frac{\gspin}{\Deltaspin} \sqrt{\nmech} \ll 1 \comma
	\label{eqn:SM:SWcondition}
\end{align}
holds, which can be satisfied by choosing a sufficiently large spin-mechanical detuning $\Deltaspin$. 
Introducing a small parameter $0 < \eta_\mathrm{SW} \ll 1$ we can rewrite Eq.~\eqref{eqn:SM:SWcondition} as 
\begin{align}	
	\nmech(\tmeas) = \eta_\mathrm{SW}^2 \frac{\Deltaspin^2}{\gspin^2} \fullstop
	\label{eqn:SM:SWconditionRewritten}
\end{align}
Since the phonon number $\nmech(\tmeas)$ increases with the probe-laser amplitude $\abs{a_\mathrm{pr,in}}$, the condition~\eqref{eqn:SM:system:SchriefferWolff} places a constraint on the maximum admissible probe-laser amplitude, 
\begin{align}
	\abs{a_\mathrm{pr,in}} \ll \abs{a_\mathrm{pr,in}^\mathrm{crit}} \comma
	\label{eqn:SM:CondAprin}
\end{align}
where our goal is to determine $\abs{a_\mathrm{pr,in}^\mathrm{crit}}$. 

In the regimes $\chi \ll \Gmech$ and $\chi \gg \Gmech$, the mechanical mode will reach its steady-state value within the measurement time $\tmeas$, i.e, we can set $\nmech(\tmeas) = \nmech^\mathrm{ss}$ and use the explicit formula for $\nmech^\mathrm{ss}$, 
\begin{align}
	\nmech^\mathrm{ss} 
		\equiv \lim_{\tau \to \infty} \nmech(\tau) 
		= \frac{4 \Gmech \Com \abs{a_\mathrm{pr,in}}^2}{\Gmech^2(1+\Com)^2 + 4 \varepsilon^2} \comma
	\label{eqn:SM:nmechss}
\end{align}
to relate $\nmech(\tmeas)$ to the probe-laser amplitude $\abs{a_\mathrm{pr,in}}$. 

However, for $\chi \gtrsim \Gmech$, a SNR of unity may be obtained before the phonon number approaches this steady-state value, i.e., $\nmech(\tmeas) < \nmech^\mathrm{ss}$. 
Introducing a second parameter $0 < \eta_\mathrm{trans} \leq 1$, we can combine all three regimes into the relation 
\begin{align}
	\nmech(\tmeas) = \eta_\mathrm{trans} \nmech^\mathrm{ss} \comma
	\label{eqn:SM:nmechtonmechss}
\end{align}
where $\eta_\mathrm{trans} < 1$ ($\eta_\mathrm{trans} = 1$) if $\chi \approx \Gmech$ ($\chi \ll \Gmech$ or $\chi \gg \Gmech$). 
Combining Eqs.~\eqref{eqn:SM:SWconditionRewritten}, \eqref{eqn:SM:nmechss}, and \eqref{eqn:SM:nmechtonmechss}, we find using $\varepsilon^2 = \chi^2$
\begin{align}
	\abs{a_\mathrm{pr,in}}^2 
	&= \frac{\Gmech^2 (1 + \Com)^2 + 4 \chi^2}{4 \Gmech \Com} \frac{\eta_\mathrm{SW}^2}{\eta_\mathrm{trans}} \left( \frac{\Deltaspin}{\gspin} \right)^2
	= \eta_\mathrm{SW}^2 \abs{a_\mathrm{pr,in}^\mathrm{crit}}^2 \comma
\end{align}
where the critical probe-laser amplitude is
\begin{align}
	\abs{a_\mathrm{pr,in}^\mathrm{crit}}^2 
	&= \frac{1}{\eta_\mathrm{trans}} \frac{\Gmech^2 (1 + \Com)^2 + 4 \chi^2}{4 \Gmech \Com} \left( \frac{\Deltaspin}{\gspin} \right)^2 \fullstop
	\label{eqn:SM:aprwithetatrans}
\end{align}

To remove the auxiliary parameter $\eta_\mathrm{trans}$ from Eq.~\eqref{eqn:SM:aprwithetatrans}, we minimize the right-hand side with respect to $\eta_\mathrm{trans}$ and (in a second step) $\Com$. 
This yields the following two lower (i.e., conservative) estimates on $\abs{a_\mathrm{pr,in}^\mathrm{crit}}$,
\begin{align}
	\abs{a_\mathrm{pr,in}^\mathrm{crit}}^2 
	\geq \left( \frac{\Deltaspin}{\gspin} \right)^2 \frac{\Gmech^2 (1 + \Com)^2 + 4 \chi^2}{4 \Gmech \Com} 
	\geq \left( \frac{\Deltaspin}{\gspin} \right)^2 \frac{\Gmech}{2} \left( \sqrt{1 + \frac{4 \chi^2}{\Gmech^2}} + 1 \right) \fullstop
	\label{eqn:SM:aprincbounds}
\end{align}
Thus, choosing $\abs{a_\mathrm{pr,in}}$ much smaller than the expressions on the right-hand side of Eq.~\eqref{eqn:SM:aprincbounds}, 
\begin{align}
	\abs{a_\mathrm{pr,in}^\mathrm{crit}}^2 
	\geq \left( \frac{\Deltaspin}{\gspin} \right)^2 \frac{\Gmech^2 (1 + \Com)^2 + 4 \chi^2}{4 \Gmech \Com} 
	\geq \left( \frac{\Deltaspin}{\gspin} \right)^2 \frac{\Gmech}{2} \left( \sqrt{1 + \frac{4 \chi^2}{\Gmech^2}} + 1 \right) 
	\gg \abs{a_\mathrm{pr,in}}^2 \comma
	\label{eqn:SM:aprincrEstimate}
\end{align}
will ensure that Eq.~\eqref{eqn:SM:SWcondition} is satisfied.
Note that the first $\geq$ sign turns into an equality for $\chi \ll \Gmech$ or $\chi \gg \Gmech$, since we then have $\eta_\mathrm{trans} = 1$. 
We choose our system parameters such that $\nmech < \ncrit$ for all considered plot values, see the inset of Fig.~\ref{fig:zerotemperature} of the main text. 
In particular, for the SiV parameters given in the main text, this amounts to $\abs{a_\mathrm{pr,in}^\mathrm{cr}}^2/\sqrt{\Gmech} \geq 30$ and $\ncrit \approx \Valuenmechcrit$, compared to an actual phonon number of $\nmech(\tmeas) = \Valuenmech$.

\subsection{Finite-temperature case}

\begin{figure}
	\centering
	\includegraphics[width=0.48\textwidth]{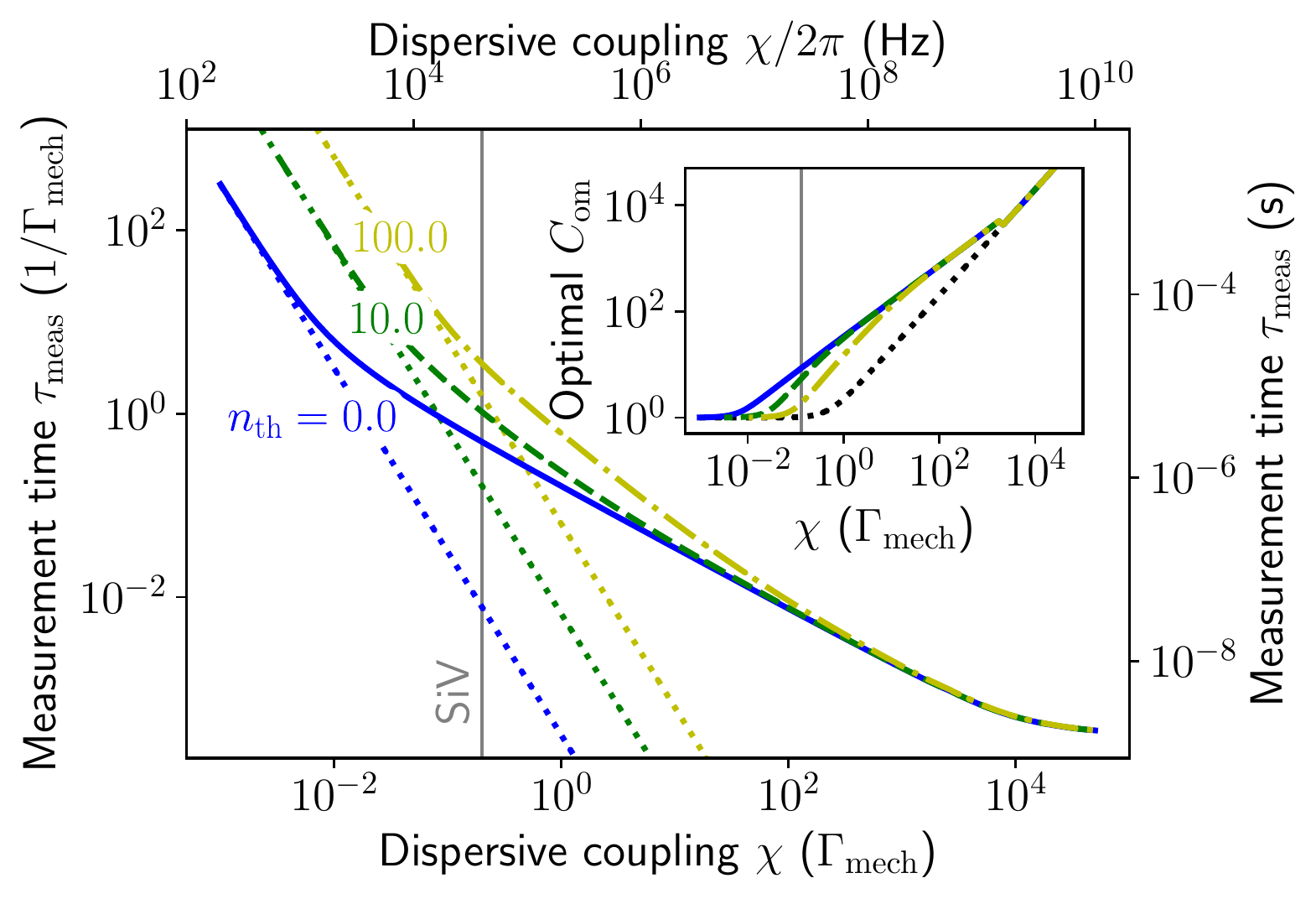}
	\caption{
		Minimum measurement time required to reach a SNR of unity in the presence of a finite-temperature mechanical bath. 
		The solid blue curve is the same as in Fig.~\ref{fig:zerotemperature} of the main text, $a_\mathrm{pr,in}/\sqrt{\Gmech} = \Valueapr$ and $\nth = 0$.
		The green (yellow) curve are for increasing thermal phonon number but fixed $a_\mathrm{pr,in}/\sqrt{\Gmech} = \Valueapr$. 
		The inset shows the optimal cooperativity $\Com$ minimizing $\tmeas$.
		As $\Com$ grows, optomechanical cooling compensates the nonzero thermal phonon number $\nth$, and the zero-temperature limit is recovered for $\chi/\Gmech \gg 1$.  
	}
	\label{fig:finitetemperature}
\end{figure}

The case of a zero-temperature environment assumed in the main text is an excellent approximation for the optical mode.
However, the mechanical mode of the OMC may have a nonzero thermal occupation $\nth$ if the temperature of the environment is not small compared to $\ommech$ or if the OMC is heated due to optical absorption. 
The SNR has then the form
\begin{align}
	\SNR^2(\tau, \nth) = \frac{\SNR^2(\tau, 0)}{1 - G(\tau)} \comma
\end{align}
where $\SNR^2(\tau,0)$ is the zero-temperature SNR given in Eq.~\eqref{eqn:SNR} of the main text and $G(\tau)$ is defined in Eq.~\eqref{eqn:SM:helperGtau}. 
The function $G(\tau)$ captures the fluctuations at finite temperature and has the limits
\begin{align}
	\lim_{\nth \to 0} G(\tau) &= 0 \comma \\
	\lim_{\tau \to \infty} G(\tau) &= - 2 \nth \comma \\
	\lim_{\chi \to \infty} G(\tau) &= 0 \fullstop
\end{align}
If the integration time is much longer than the mechanical ringup time, $\tmeas \gg 1/\Gmech$, $G(\tau)$ has decayed to its steady-state value such that the noise is purely diffusive and simply enhanced by a factor $\sqrt{1 + 2 \nth}$, i.e., $\mathrm{N}(\tau) = \sqrt{2 \kappa \tau (1 + 2 \nth)}$. 
Thus, for $\chi \ll \Gmech$, the impedance-matching condition $\Com = 1$ will be unchanged and the measurement time is 
\begin{align}
	\tmeas \to \frac{\Gmech^2}{8 \abs{a_\mathrm{pr,in}}^2 \chi^2} (1 + 2 \nth) \comma
\end{align}
which generalizes Eq.~\eqref{eqn:tmeasWeakCoupling} of the main text. 
For $\chi \gtrsim \Gmech$, the transient dynamics becomes relevant and $G(\tau)$ depends explicitly on the integration time. 
However, the optimal cooperativity in this regime increases, $\Com > 1$, which leads to an increased optomechanical cooling and effectively reduces $\nth$. 
For $\Com \gg 1, \chi/\Gmech$, one can show that
\begin{align}
	G(\tau) \approx 8 \frac{\nth}{\Com} \frac{2 - x - 2 e^{-x/2}}{x} \comma 
\end{align}
where $x = \Gmech \Com \tau$. 
Since $[2 - x - 2 e^{-x/2}]/x \in [-1,0]$, optomechanical cooling will dominate if $\Com \gtrsim 10 \nth$, such that we recover the zero-temperature result $G(\tau) \to 0$.
In particular, the fundamental limit deep in the strong-dispersive-coupling regime, $\chi/\Gmech \gg 1, \nth$, 
\begin{align}
	\tmeas \to \frac{1}{8 \abs{a_\mathrm{pr,in}}^2} \comma
\end{align}
is the same as for the zero-temperature case, see Eq.~\eqref{eqn:FundamentalLimittmeas} of the main text. 
This crossover to a zero-temperature result is illustrated in Fig.~\ref{fig:finitetemperature}, where we plot the minimum measurement time and the corresponding optimal cooperativity for different thermal phonon numbers $\nth$. 
Interestingly, the range of $\chi$ values for which $\Com \approx 1$ grows with $\nth$.
This means that, for small $\chi \ll \Gmech$, it is more beneficial to maintain the impedance-matching condition $\Com = 1$ (which maximizes the output signal) than to increase $\Com$ (which would enhance the optomechanical cooling at the expense of a reduced signal).

\subsection{Bounded optomechanical cooperativity}

\begin{figure*}
	\centering
	\subfigure[]{
		\includegraphics[width=0.48\textwidth]{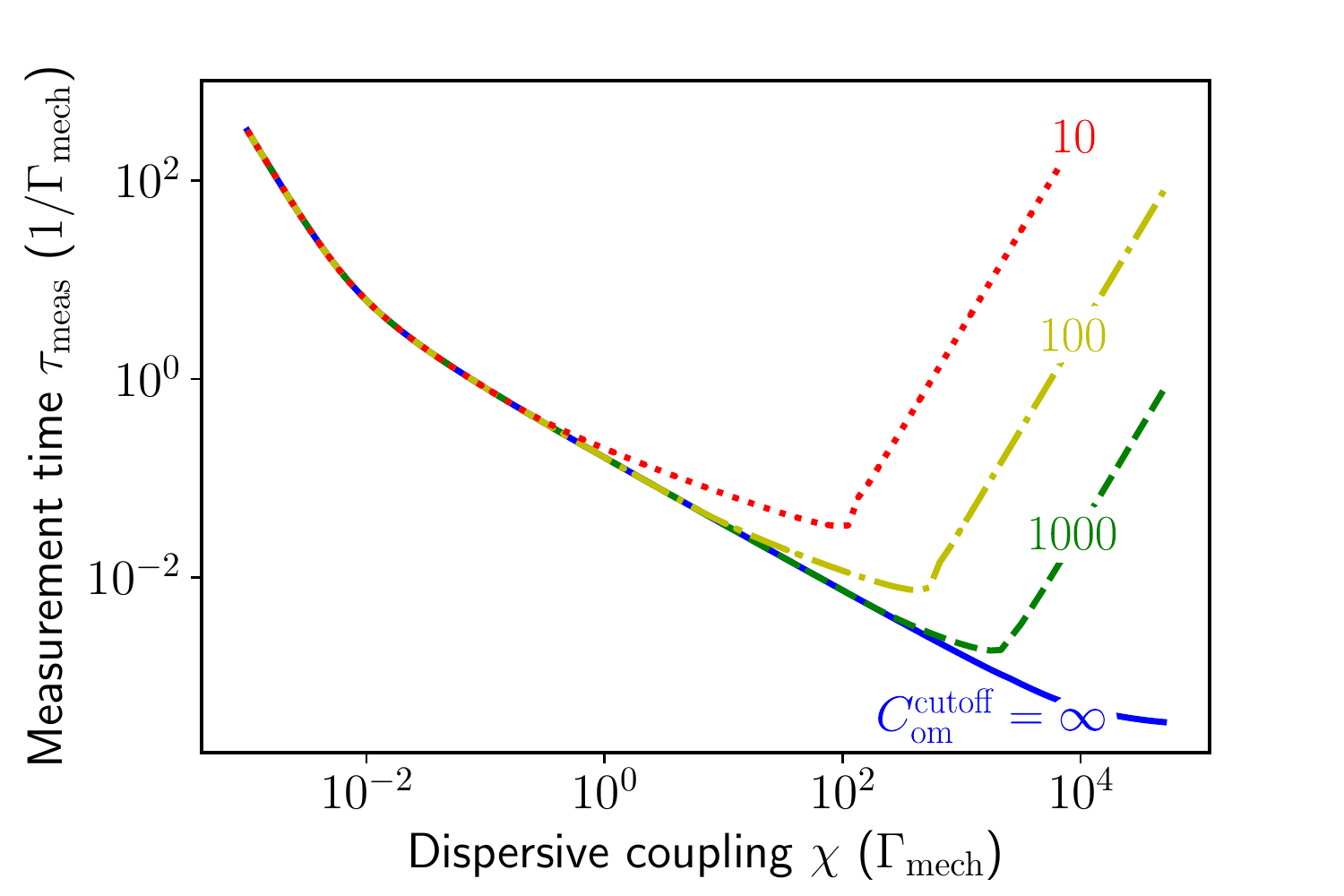}
	}
	\subfigure[]{
		\includegraphics[width=0.48\textwidth]{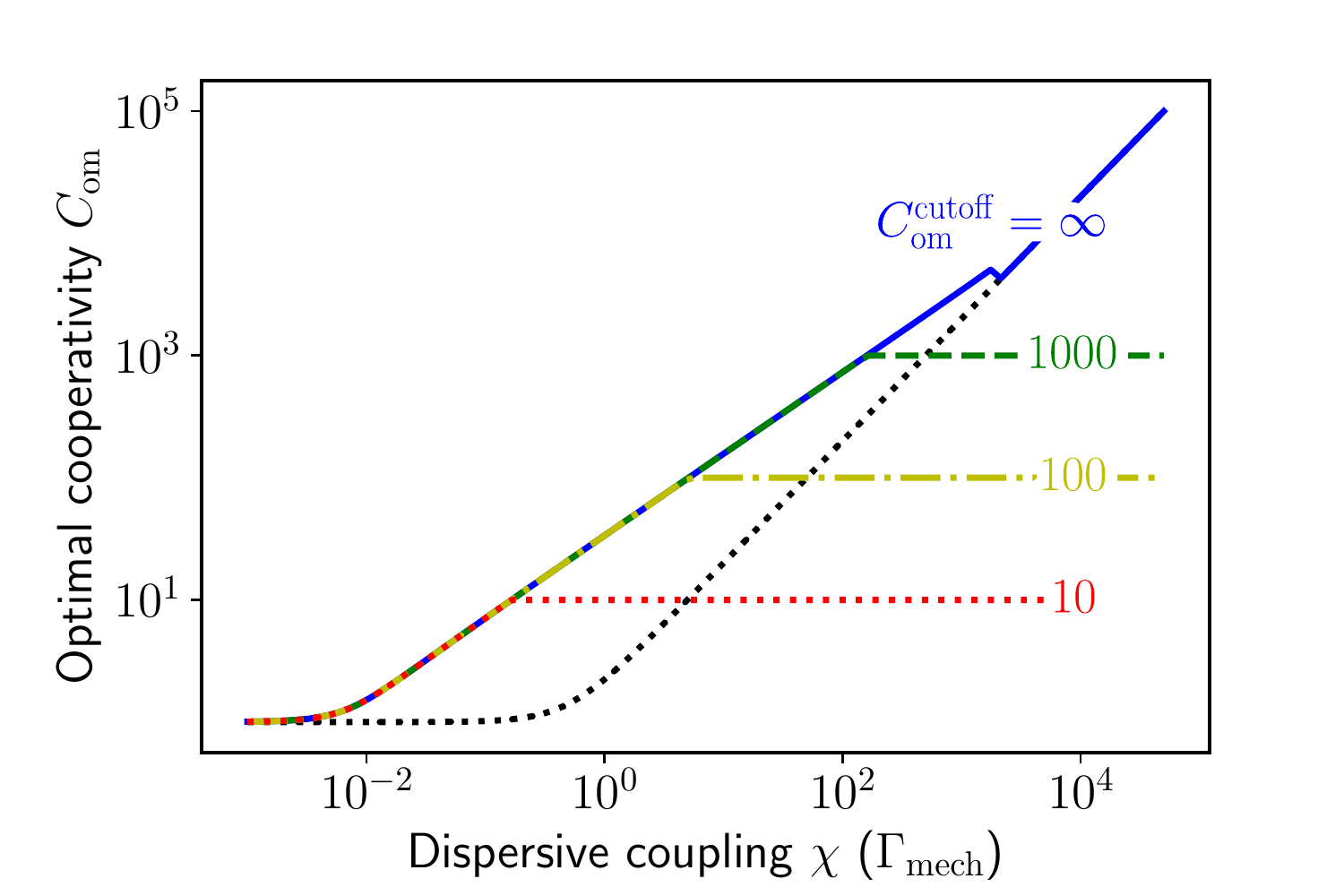}
	}
	\subfigure[]{
		\includegraphics[width=0.48\textwidth]{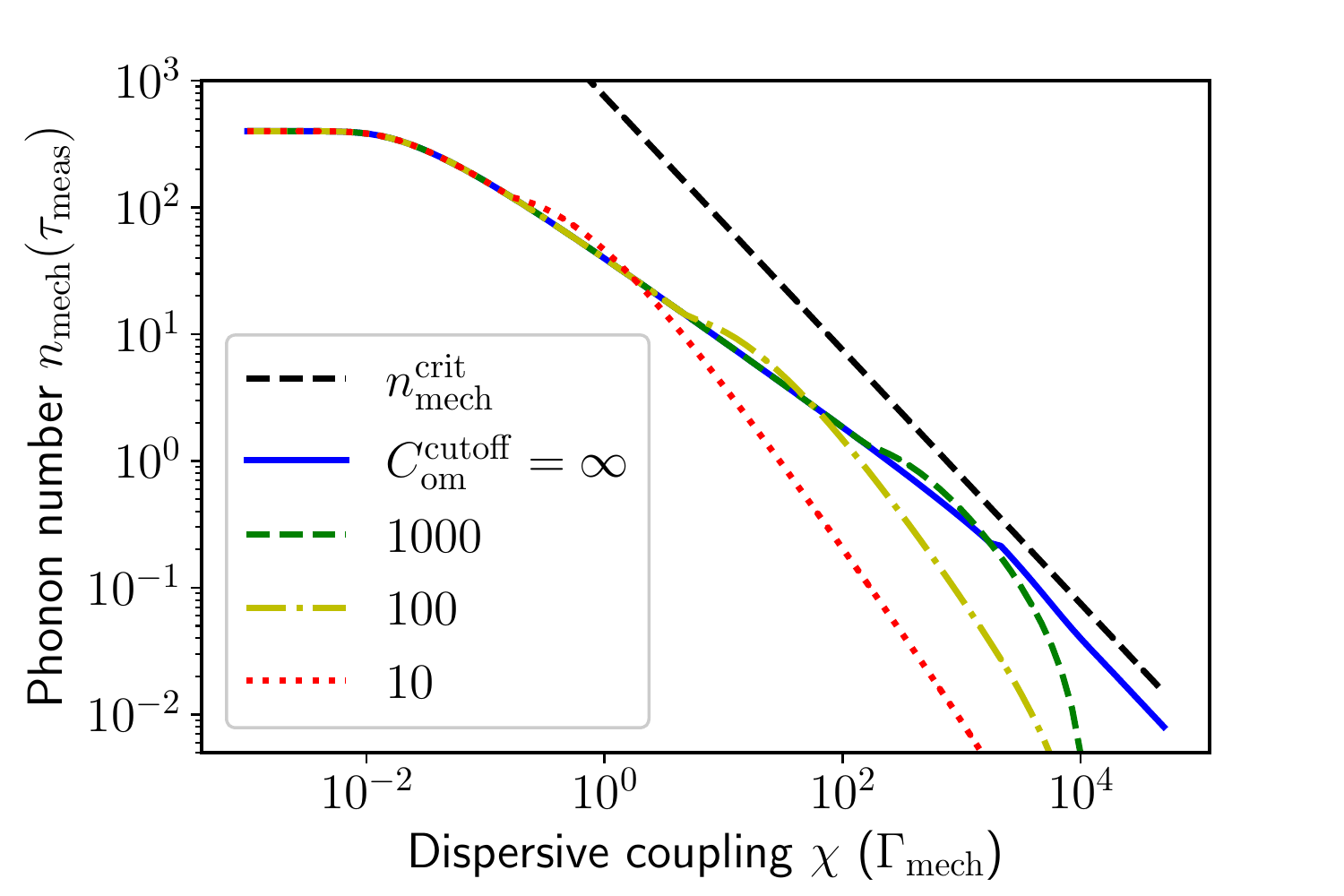}
	}
	\subfigure[]{
		\includegraphics[width=0.48\textwidth]{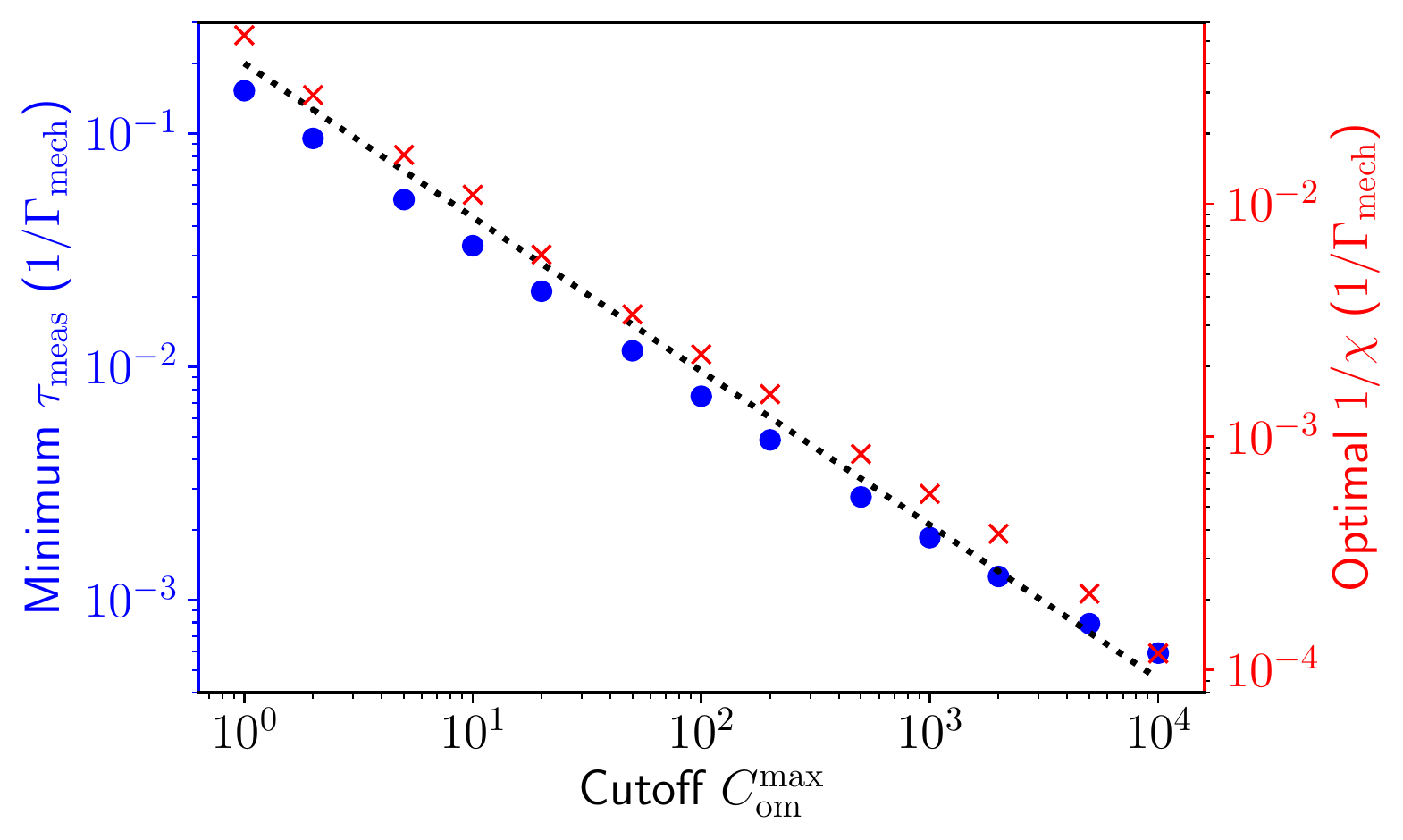}
	}
	\caption{
		Single-spin readout with a constrained optomechanical cooperativity. 
		(a) Minimum measurement time $\tmeas$ as a function of the dispersive coupling strength. 
		The blue curve shows the data of Fig.~\ref{fig:zerotemperature} of the main text for $a_\mathrm{pr,in}/\sqrt{\Gmech} = \Valueapr$ and $\nth = 0$.
		For all curves, the cooperativity $\Com$ has been optimized with a constraint $\Com \leq \Com^\mathrm{cutoff}$ indicated by the labels. 
		(b) Corresponding optimal cooperativity.
		The dotted black line indicates the optimal cooperativity $\Com = \sqrt{1 + 4 \chi^2/\Gmech^2}$ if transient dynamics can be ignored. 
		(c) Corresponding mechanical phonon number $\nmech(\tmeas)$ as a function of $\chi$. 
		Like in the inset of Fig.~\ref{fig:zerotemperature} of the main text, the black dashed curve indicating the critical phonon number has been calculated by varying $\gspin$ while keeping the spin-mechanical detuning $\Deltaspin/\Gmech = \ValueDeltasmOverGamma$ fixed, and using $\kappa/\Gmech = \ValueRatiokappaoverGamma$.
		(d) Global minimum of the measurement times as a function of $\chi$ shown in (a) (blue dots, left axis) and the corresponding values of $\chi$ (red crosses, right axis). 
		The dotted black line indicates a $1/\Com^{2/3}$ scaling with prefactor $0.2$ (left axis) and $0.04$ (right axis).
	}
	\label{fig:SM:Cooperativity_cutoff}
\end{figure*}

The discussion in the main text suggest that the optimal cooperativity diverges with increasing dispersive coupling, $\Com \to 2 \chi/\Gmech$. 
However, such a scaling is not realistic for the following two reasons. 
First, from an experimental point of view, the optical cooperativity will be bounded from above by the available laser power as well as the maximum heating rate tolerable by the optomechanical crystal and the optical components in the setup. 
Second, from a theoretical point of view, correction terms to the derivation in Sec.~\ref{sec:SM:OutputFields} will become relevant if the optically-enhanced coupling strength $G$ approaches the strong-coupling regime $G \gtrsim \kappa$ \cite{Lemonde2013}. 
For the chosen parameters, this will occur if $\Com \gtrsim 4 \times 10^4$. 
Therefore, we also provide simulations of $\tmeas$, $\Com$, and $\nmech(\tmeas)$ with different cutoffs on $\Com$, which are shown in Fig.~\ref{fig:SM:Cooperativity_cutoff}.

As soon as the cooperativity reaches the cutoff value, the decrease of $\tmeas$ with increasing $\chi/\Gmech$ slows down, followed by a sharp increase of $\tmeas$ for even larger values of the dispersive coupling [see Fig.~\ref{fig:SM:Cooperativity_cutoff}(a)]. 
At the same time, $\nmech(\tmeas)$ decreases below its value in the case of no cutoff [see Fig.~\ref{fig:SM:Cooperativity_cutoff}(c)]. 
These effects can be understood by the following intuitive picture. 
In the transient regime, $\Com$ is increased beyond unity to increase the mechanical damping rate and thus to speed up the mechanical response to the weak probe laser.
A bound on the cooperativity prevents reaching the optimal value of the enhanced mechanical damping rate but still leads to a reduction of the overall measurement time $\tmeas$ compared to the case with $\Com = 1$. 
At the same time, the increased mechanical damping rate ensures that the mechanical mode (whose resonance frequency is shifted to $\ommech \pm \chi$ depending on the spin state) can still be driven by the optical probe laser at frequency $\ommech$. 
If the dispersive coupling is increased significantly beyond $\chi \approx \Gmech \Com^\mathrm{cutoff}$, the mechanical mode is too off-resonant to be driven by the probe laser, $\nmech$ drops, and the measurement time increases strongly.

In Fig.~\ref{fig:SM:Cooperativity_cutoff}(d), we show the minimum achievable measurement times at the kink of the curves in Fig.~\ref{fig:SM:Cooperativity_cutoff}(a), and the corresponding optimal values of the dispersive coupling strength.

\section{Relevant solid-state defects}
\subsection{General criteria}
\label{sec:SM:GeneralCriteria}
As stressed in the main text, the OMIT scheme is not restricted to a particular type of solid-state defect. 
The basic conditions a potential solid-state defect should satisfy such that our OMIT readout protocol is feasible, are: 
\begin{itemize}
	\item a large strain susceptibility (i.e., change of the energy levels per strain),
	\item sufficiently long relaxation times to enable QND readout, 
	\item sufficiently long coherence times for quantum sensing applications, and
	\item the ability to embed it into an optomechanical system.
\end{itemize}
Here, we make these intuitive criteria more quantitative and derive the two feasibility conditions stated in the main text. 
Providing specific numbers for a broad variety of spin defects is close to impossible since, to the best of our knowledge, the spin-strain coupling $\gspin$ has not been measured or estimated for most types of defects. 
However, the general conditions we derive can easily be evaluated once these measurements have been done.
The corresponding OMIT readout fidelity and measurement time should then be compared to other readout techniques for the solid-state defect of interest. 
As discussed in the main text, some solid-state defects can be easily read out optically, but optical readout of many other defects is very unwieldy or even impossible.
This opens a wide range of applications for strain-mediated OMIT readout.

First, we note that, for QND readout, dephasing of the spins during the readout process is irrelevant since it does not change the expectation value $\cerw{\hat{\sigma}_z}$. 
Thus, the relevant figure of merit is the spin relaxation time [which may stem from single-spin relaxation time $T_1$ or the Purcell decay term in Eq.~\eqref{eqn:SM:system:QME}], compared to the readout time $\tau_\mathrm{meas}$, 
\begin{align}
	\tmeas \ll \min(T_1, \tau_\mathrm{Purcell}) \fullstop
	\label{eqn:SM:Criterion:RelaxatinRates}
\end{align}
In the following, we assume that the spin-strain coupling is rather weak, $\chi \ll \Gmech$, which we expect to be the generic situation for most defects (e.g., NV centers in diamond and divacancy defects in SiC; see Sec.~\ref{sec:SM:ProspectsOtherDefects} below). 
Since a small strain coupling $\gspin$ can be collectively enhanced in a large ensemble of spins, we will consider $N$ spins in the following. 
In this case, the measurement time is given by
\begin{align}
	\tmeas = \frac{\Gmech}{8 \chi^2 N^2 \nmech} \fullstop
\end{align} 
One may think that a small dispersive coupling $N \chi$ can always be compensated using more probe phonons. 
However, as discussed in Sec.~\ref{sec:SM:MaximumMechanicalPhononNumber}, correction terms to the dispersive spin-mechanical interaction define a critical phonon number. 
We assume that we always chose the number of probe phonons as large as possible without exceeding the critical phonon number, 
\begin{align}
	\frac{\chi N \nmech}{\Deltaspin} = \frac{\gspin^2 N \nmech}{\Deltaspin^2} \stackrel{!}{=} \eta_\mathrm{SW}^2 \ll 1 \fullstop
\end{align}
With increasing spin-mechanical detuning $\Deltaspin$, one can thus choose a larger number of probe phonons, 
\begin{align}
	\nmech &= \eta_\mathrm{SW}^2 \frac{\Deltaspin^2}{N \gspin^2} \comma
\end{align}
and the measurement time depends only on $\Gmech$ and the collectively enhanced strain coupling,
\begin{align}
	\tau_\mathrm{meas} 
	= \frac{\Gmech}{8 \gspin^2 N} \fullstop
\end{align}
From Eq.~\eqref{eqn:SM:system:QME}, we see that the Purcell decay time is given by
\begin{align}
	\tau_\mathrm{Purcell} = \frac{\Deltaspin}{\Gmech \chi N^\beta} = \frac{\Deltaspin^2}{\Gmech \gspin^2 N^\beta} \comma
\end{align}
where the exponent $\beta$ captures the fact that the collective decay rate depends on the polarization of the initial state. 
If the spins are highly polarized along the $z$ axis, we have $\beta = 1$, but if each spin has $\cerw{\sigma_z} \approx 0$, we have $\beta = 2$.

If Eq.~\eqref{eqn:SM:Criterion:RelaxatinRates} is limited by Purcell decay, $\tau_\mathrm{Purcell} < T_1$, one can increase $\Deltaspin$ to increase $\tau_\mathrm{Purcell}$ quadratically, while $\tmeas$ remains constant since we are also allowed to use more probe phonons, $n_\mathrm{mech} \propto \Delta_\mathrm{sm}^2$.
A QND measurement is achieved if the detuning satisfies 
\begin{align}
	\frac{\Deltaspin^2}{\Gmech^2} \gg \frac{N^{\beta - 1}}{8} \fullstop
	\label{eqn:SM:Criterion:Detuning}
\end{align}
Importantly, the corresponding bound on the phonon number does not increase with the ensemble size $N$ (and in fact decreases for $\beta = 1$), 
\begin{align}
	\nmech \gg \frac{\eta_\mathrm{SW}^2 \Gmech^2}{8 \gspin^2} N^{\beta - 2} \fullstop
\end{align}

When Purcell decay is suppressed by increasing $\Delta_\mathrm{sm}$, Eq.~\eqref{eqn:SM:Criterion:RelaxatinRates} will ultimately become limited by the intrinsic relaxation, $T_1 < \tau_\mathrm{Purcell}$. 
We then have to satisfy the condition $\tau_\mathrm{meas} \ll T_1$, which is equivalent to the collective cooperativity criterion
\begin{align}
	\frac{4 \gspin^2 N}{\Gmech \gamma_\mathrm{rel}} \gg \frac{1}{2} \comma
	\label{eqn:SM:Criterion:CollectiveCooperativity}
\end{align}
where $\gamma_\mathrm{rel} = 2 \pi/T_1$. 
A weak spin-strain coupling can be collectively enhanced using a large ensemble of spin defects, until this condition is satisfied.

\subsection{Estimated strain coupling for SiV defects}
\label{sec:SM:StrainCoupling}

In this section, we estimate the strain coupling of an SiV defect in the presence of a magnetic field. 
The negatively charged SiV center is an interstitial point defect in the diamond lattice where two carbon atoms have been replaced by a silicon atom.
We chose the axis between the missing carbon atoms to be the $z$ axis. 
The silicon atom is placed along the $z$ axis in the middle between the two missing carbon atoms, such that the entire defect has an inversion symmetry about the position of the silicon atom. 
Therefore, the SiV defect belongs to the $D_{3d}$ point group, and its electronic orbitals have $A$ or $E$ symmetry and can have even ($g$) or odd ($u$) parity with respect to inversion about the Si position \cite{Gali2013,Hepp2014}. 
In the ground (excited) state, an unpaired hole is in the $e_{gx},e_{gy}$ ($e_{ux},e_{uy}$) orbitals, each of which is twofold degenerate due to the spin degree of freedom. 
The resulting four-fold degeneracy of the ground-state and excited-state manifolds is partially lifted by the spin-orbit (SO) interaction, which splits the ground-state (excited-state) manifold into two spin-degenerate doublets separated by a spin-orbit splitting $\lambda_{\mathrm{SO},g} \approx 46\,\mathrm{GHz}$ ($\lambda_{\mathrm{SO},u} \approx 255\,\mathrm{GHz}$).
For  each manifold $\alpha \in \{g,u\}$, the spin-orbit Hamiltonian is
\begin{align}
	\hat{H}_{\mathrm{SO},\alpha} = - \frac{\lambda_{\mathrm{SO},\alpha}}{2} \hat{L}_z^{(\alpha)} \otimes \hat{S}_z \comma
\end{align}
where the $z$ component of the orbital angular momentum operator in the basis $\{ \ket{e_{\alpha x}}, \ket{e_{\alpha,y}} \}$ is \cite{Hepp2014PRL}
\begin{align}
	\hat{L}_z^{(\alpha)} = \begin{pmatrix}
		0 & -i \\
		i & 0 
	\end{pmatrix} \comma
\end{align}
and the $z$ component of the spin operator is $\hat{S}_z = \ket{\uparrow} \bra{\uparrow} - \ket{\downarrow}\bra{\downarrow}$. 
In the absence of strain or magnetic fields, $\hat{H}_{\mathrm{SO},\alpha}$ is diagonal in the basis $\{ \ket{e_{\alpha -} \downarrow}, \ket{e_{\alpha +} \uparrow}, \ket{e_{\alpha +} \downarrow}, \ket{e_{\alpha -} \uparrow} \}$ with eigenvalues $\{- \lambda_{\mathrm{SO},\alpha}/2, - \lambda_{\mathrm{SO},\alpha}/2, + \lambda_{\mathrm{SO},\alpha}/2, + \lambda_{\mathrm{SO},\alpha}/2\}$.
The orbital eigenstates are defined as $\ket{e_{\alpha \pm}} = \mp (\ket{e_{\alpha x}} \pm i \ket{e_{\alpha y}})/\sqrt{2}$, where the subscripts $\pm$ denote the orbital-angular-momentum projection of the states, $\hat{L}_z^{(\alpha)} \ket{e_{\alpha \pm}} = \pm \ket{e_{\alpha \pm}}$.

The remaining degeneracies are lifted by magnetic fields and strain, which are modeled by the following Hamiltonians \cite{Hepp2014PRL,Meesala2018}, 
\begin{align}
	\hat{H}_{\mathrm{Z},\alpha} &= \gamma_\mathrm{L} B_z \hat{L}_z^{(\alpha)} \otimes \hat{\1} + \gamma_\mathrm{S} \hat{\1}^{(\alpha)} \otimes (B_x \hat{S}_x + B_y \hat{S}_y + B_z \hat{S}_z) \comma \\
	\hat{H}_{\mathrm{strain},\alpha} &= \left[ \varepsilon_{A_{1g}}^\alpha \left( \ket{e_{\alpha x}} \bra{e_{\alpha x}} + \ket{e_{\alpha y}} \bra{e_{\alpha y}} \right) + \varepsilon_{E_{g x}}^\alpha \left( \ket{e_{\alpha x}} \bra{e_{\alpha x}} - \ket{e_{\alpha y}} \bra{e_{\alpha y}} \right) + \varepsilon_{E_{g y}}^\alpha \left( \ket{e_{\alpha x}} \bra{e_{\alpha y}} + \ket{e_{\alpha y}} \bra{e_{\alpha x}} \right) \right] \otimes \hat{\1} \comma
\end{align}
where $\hat{S}_x = \ket{\uparrow}\bra{\downarrow} + \ket{\downarrow}\bra{\uparrow}$ and $\hat{S}_y = (\ket{\uparrow} \bra{\downarrow} - \ket{\downarrow}\bra{\uparrow})/i$. 
The energies $\varepsilon_{A_{1g}}^\alpha$, $\varepsilon_{E_{gx}}^\alpha$, and $\varepsilon_{E_{gy}}^\alpha$ depend on the strain tensor decomposed in terms of the irreducible representations of the $D_{3d}$ point group, and the associated strain susceptibilities of the $\alpha \in \{g, u\}$ manifolds, 
\begin{align}
	\varepsilon_{A_{1g}}^\alpha &= t_\perp^\alpha (\epsilon_{xx} + \epsilon_{yy}) + t_\parallel^\alpha \epsilon_{zz} \comma \\
	\varepsilon_{E_{gx}}^\alpha &= d^\alpha (\epsilon_{xx} - \epsilon_{yy}) + f^\alpha \epsilon_{zx} \comma \\
	\varepsilon_{E_{gy}}^\alpha &= - 2 d^\alpha \epsilon_{xy} + f^\alpha \epsilon_{yz} \comma
\end{align}
where the strain susceptibilities are $d^g = 1.3 \,\mathrm{PHz}/\mathrm{strain}$ as well as $f^g = - 1.7\,\mathrm{PHz}/\mathrm{strain}$, and the gyromagnetic ratios are $\gamma_\mathrm{L} = 0.1 \times 14\,\mathrm{GHz}/\mathrm{T}$ and $\gamma_\mathrm{S} = 14\,\mathrm{GHz}/\mathrm{T}$ \cite{Meesala2018}. 
The difference of the strain susceptibilities $t_\parallel^\alpha$ and $t_\perp^\alpha$ for the $g$ and $e$ manifolds have been measured in Ref.~\onlinecite{Meesala2018}, too, but they are not important here since the $\varepsilon_{A_{1g}}^\alpha$ terms will be irrelevant in the following analysis.

The total Hamiltonian in the $\alpha$ manifold, expressed in the SO eigenbasis $\{ \ket{e_{\alpha -} \downarrow}, \ket{e_{\alpha +} \uparrow}, \ket{e_{\alpha +} \downarrow}, \ket{e_{\alpha -} \uparrow} \}$ is thus
\begin{align}
	\hat{H}_\alpha = \begin{pmatrix}
		- \frac{\lambda_{\mathrm{SO},\alpha}}{2} - (\gamma_\mathrm{S} + \gamma_\mathrm{L}) B_z & 
		0 & 
		- \varepsilon_{E_{gx}}^\alpha - i \varepsilon_{E_{gy}}^\alpha & 
		\gamma_\mathrm{S} (B_x + i B_y) \\
		0 & 
		- \frac{\lambda_{\mathrm{SO},\alpha}}{2} + (\gamma_\mathrm{S} + \gamma_\mathrm{L}) B_z &
		\gamma_\mathrm{S} (B_x - i B_y) & 
		- \varepsilon_{E_{gx}}^\alpha + i \varepsilon_{E_{gy}}^\alpha \\
		- \varepsilon_{E_{gx}}^\alpha + i \varepsilon_{E_{gy}}^\alpha & 
		\gamma_\mathrm{S}(B_x + i B_y) & 
		+ \frac{\lambda_{\mathrm{SO},\alpha}}{2} - (\gamma_\mathrm{S} - \gamma_\mathrm{L}) B_z & 
		0 \\
		\gamma_\mathrm{S} (B_x - i B_y) & 
		- \varepsilon_{E_{gx}}^\alpha - i \varepsilon_{E_{gy}}^\alpha & 
		0 & 
		+ \frac{\lambda_{\mathrm{SO},\alpha}}{2} + (\gamma_\mathrm{S} - \gamma_\mathrm{L}) B_z\\
	\end{pmatrix} + \varepsilon_{A_{1g}}^\alpha \hat{\1}_{4\times4} \fullstop
\end{align}

Our goal is to identify an effective two-level system within the ground-state manifold of $\hat{H}_\alpha$. 
We therefore set $\alpha = g$ in the following and suppress this subscript for simplicity. 
We also ignore the $\varepsilon_{A_{1g}}^\alpha$ term, which only contributes a constant energy shift of the $g$ and $u$ manifolds, and we set $B_y = 0$. 
Both the Zeeman and the strain terms mix different SO eigenstates. 
However, for SiV defects in a diamond OMC, the Zeeman terms will be of the order of $\mathrm{GHz}$ (i.e., comparable to the SO splitting) whereas the strain terms will be of the order of $\mathrm{MHz}$. 
We therefore diagonalize $\hat{H}_\mathrm{SO} + \hat{H}_\mathrm{Z}$ and treat the strain terms $\hat{H}_\mathrm{strain}$ perturbatively.
The eigenvectors of $\hat{H}_\mathrm{SO} + \hat{H}_\mathrm{Z}$ will be denoted by $\ket{e_\tau \sigma}'$ with $\tau \in \{+,-\}$ and $\sigma \in \{\uparrow,\downarrow\}$, where $\ket{e_{\tau} \sigma}$ is the corresponding eigenstate of $\hat{H}_\mathrm{SO}$ to which $\ket{e_\tau \sigma}'$ reduces in the limit of vanishing magnetic field.
The associated energies are
\begin{align}
	E_{\tau,\sigma} = \tau \left[ \gamma_\mathrm{L} B_z - \frac{2 \delta_{\sigma,\uparrow} - 1}{2} \sqrt{4 \gamma_\mathrm{S}^2 B_x^2 + (\lambda_\mathrm{SO} - 2 \tau \gamma_\mathrm{S} B_z)^2 } \right] \comma
\end{align}
where $\delta_{\sigma,\sigma'}$ denotes the Kronecker delta. 
A purely off-axis magnetic field, $B_z = 0$ but $B_x \neq 0$, simply shifts the $\ket{e_- \downarrow}'$ and $\ket{e_+ \uparrow}'$ levels with respect to the $\ket{e_+ \downarrow}'$ and $\ket{e_- \uparrow}'$ levels, but does not lift their respective degeneracy. 
We thus need a finite $B_z$ component, too, to define an effective two-level system using the $\ket{e_- \downarrow}'$ and $\ket{e_+ \uparrow}'$ levels. 
This two-level system should maintain a fixed detuning from the mechanical mode of the OMC, i.e., $E_{+,\uparrow} - E_{-,\downarrow} = \omspin = \ommech - \Deltaspin$ should be independent of the chosen magnetic field.
For $\omspin/[2(\gamma_\mathrm{L} + \gamma_\mathrm{S})] \leq B_z \leq \omspin/(2 \gamma_\mathrm{L})$, this condition can be satisfied by choosing the $B_x$ component as
\begin{align}
	\abs{B_x} 
	&= \frac{\sqrt{(2 \gamma_\mathrm{S} B_z)^2 - (2 \gamma_\mathrm{L} B_z - \omspin)^2} \sqrt{\lambda_\mathrm{SO}^2 - (2 \gamma_\mathrm{L} B_z - \omspin)^2}}{2 \gamma_\mathrm{S} \abs{2 \gamma_\mathrm{L} B_z - \omspin}} \fullstop
	\label{eqn:SM:SiVTuningBx}
\end{align}
For $B_z \gtrsim \omspin/[(2 \gamma_\mathrm{L} + \gamma_\mathrm{S})]$, $\gamma_\mathrm{S} \gg \gamma_\mathrm{L}$, and $\lambda_\mathrm{SO} \gg \omspin$, the off-axis magnetic field scales approximately as 
\begin{align}
	\abs{B_x} &\approx \frac{\lambda_\mathrm{SO}}{\sqrt{\gamma_\mathrm{S} \omspin}} \sqrt{B_z - \frac{\omspin}{2 \gamma_\mathrm{S}}} \fullstop
\end{align}

The $B_x$ field increases very quickly with growing $B_z$, which can be understood by the following intuitive argument. 
For simplicity, we consider the limit $\gamma_\mathrm{L} \to 0$, i.e., the Zeeman splitting of the high-energy ($\ket{e_{\alpha+} \downarrow}$ and $\ket{e_{\alpha-} \uparrow}$, at energy $\lambda_\mathrm{SO}/2$) and low-energy ($\ket{e_{\alpha-} \downarrow}$ and $\ket{e_{\alpha+} \uparrow}$, at energy $-\lambda_\mathrm{SO}/2$) spin-orbit doublet is identical. 
The minimum $B_z$ field which yields a real solution for $B_x$ is $B_z = \omspin/2 \gamma_\mathrm{S}$, i.e., the magnetic field parallel to the SiV axis splits the doublets just enough to generate the desired level splitting $\omspin$. 
For a $7\,\mathrm{GHz}$ transition frequency, this happens at $B_z = 0.5\,\mathrm{T}$. 
If the $B_z$ field is further increased, the transition frequency $E_{+,\uparrow} - E_{-,\downarrow}$ is larger than $\omspin$ and one needs an off-axis magnetic field $B_x$ to compensate the mismatch. 
This is possible because the $B_x$ field causes separate avoided level crossings between the the outer ($\ket{e_{\alpha-} \uparrow}$ and $\ket{e_{\alpha-} \downarrow}$) and inner ($\ket{e_{\alpha+} \downarrow}$ and $\ket{e_{\alpha+} \uparrow}$) Zeeman-split states. 
For $\gamma_\mathrm{S} B_x \gg \lambda_\mathrm{SO}$, the upper and lower Zeeman-split states will converge to the energies $\gamma_\mathrm{S} B_x$ and $-\gamma_\mathrm{S} B_x$, respectively.
Therefore, the transition frequencies between the two high-energy states (and, likewise, between the two low-energy states) must ultimately vanish, and the excess detuning caused by the $B_z$ field will be canceled for some intermediate $B_x$ field.
The convergence of the transition frequencies will occur if $\gamma_\mathrm{S} B_x \gtrsim \lambda_\mathrm{SO}$, which yields $B_x \gtrsim 3.3\,\mathrm{T} \gg B_z$.

We now investigate how the perturbation given by the strain Hamiltonian $\hat{H}_\mathrm{strain}$ affects the eigenstates $\{ \ket{e_{\alpha -} \downarrow}', \ket{e_{\alpha +} \uparrow}', \ket{e_{\alpha +} \downarrow}', \ket{e_{\alpha -} \uparrow}' \}$
of $\hat{H}_\mathrm{SO} + \hat{H}_\mathrm{Z}$. 
The magnetic field components perpendicular to the $z$ axis mix SO eigenstates with the same orbital projection $\tau$, i.e., $\ket{e_- \downarrow}' = c_{-\downarrow\downarrow} \ket{e_- \downarrow} + c_{-\downarrow\uparrow} \ket{e_- \uparrow}$, $\ket{e_+ \uparrow}' = c_{+\uparrow\uparrow} \ket{e_+ \uparrow} + c_{+\uparrow\downarrow} \ket{e_+ \downarrow}$, etc.
In contrast, the strain Hamiltonian mixes SO eigenstates with the same spin projection $\sigma$. 
Therefore, the strain Hamiltonian rewritten in the basis $\{ \ket{e_{\alpha -} \downarrow}', \ket{e_{\alpha +} \uparrow}', \ket{e_{\alpha +} \downarrow}', \ket{e_{\alpha -} \uparrow}' \}$ has the form
\begin{align}
	\hat{H}_\mathrm{strain}' = \begin{pmatrix}
		0 & c_1 (\varepsilon_{E_{gx}} + i \varepsilon_{E_{gy}}) & c_2 (\varepsilon_{E_{gx}} + i \varepsilon_{E_{gy}}) & 0 \\
		c_1^* (\varepsilon_{E_{gx}} - i \varepsilon_{E_{gy}}) & 0 & 0 & c_3 (\varepsilon_{E_{gx}} - i \varepsilon_{E_{gy}}) \\
		c_2^* (\varepsilon_{E_{gx}} - i \varepsilon_{E_{gy}}) & 0 & 0 & c_4 (\varepsilon_{E_{gx}} - i \varepsilon_{e_{gy}}) \\
		0 & c_3^* (\varepsilon_{E_{gx}} + i \varepsilon_{E_{gy}}) & c_4^* (\varepsilon_{E_{gx}} + i \varepsilon_{e_{gy}}) & 0
	\end{pmatrix} \comma
\end{align}
where the parameters $c_1$ to $c_4$ depend on the expansion coefficients $c_{\tau \sigma \sigma'}$. 
The terms proportional to $c_1$ couple the two levels $\ket{e_- \downarrow}'$ and $\ket{e_+ \uparrow}'$ of the effective two-level system to the mechanical mode, and they represent the desired strain coupling term. 
However, the $c_2$ and $c_3$ terms also introduce undesired couplings to the other states $\ket{e_+ \downarrow}'$ and $\ket{e_- \uparrow}'$. 
The energy gap between the upper two-level-system state $\ket{e_+ \uparrow}'$ and the lower state $\ket{e_+ \downarrow}'$ of the remaining states grows with $B_z$ and is at least $\lambda_\mathrm{SO} - \gamma_\mathrm{S} \omspin/(\gamma_\mathrm{L} + \gamma_\mathrm{S}) \approx \lambda_\mathrm{SO} \gg \omspin$.
These transitions are thus highly off-resonant and can be ignored in a rotating wave approximation. 
We therefore obtain a block-diagonal structure of $\hat{H}_\mathrm{strain}'$ with a well-defined two-level system formed by $\{ \ket{e_- \downarrow}', \ket{e_+ \uparrow}' \}$, which is coupled to a mechanical mode with a magnetic-field-tunable coupling strength.

\begin{figure}
	\centering
	\includegraphics[width=0.48\textwidth]{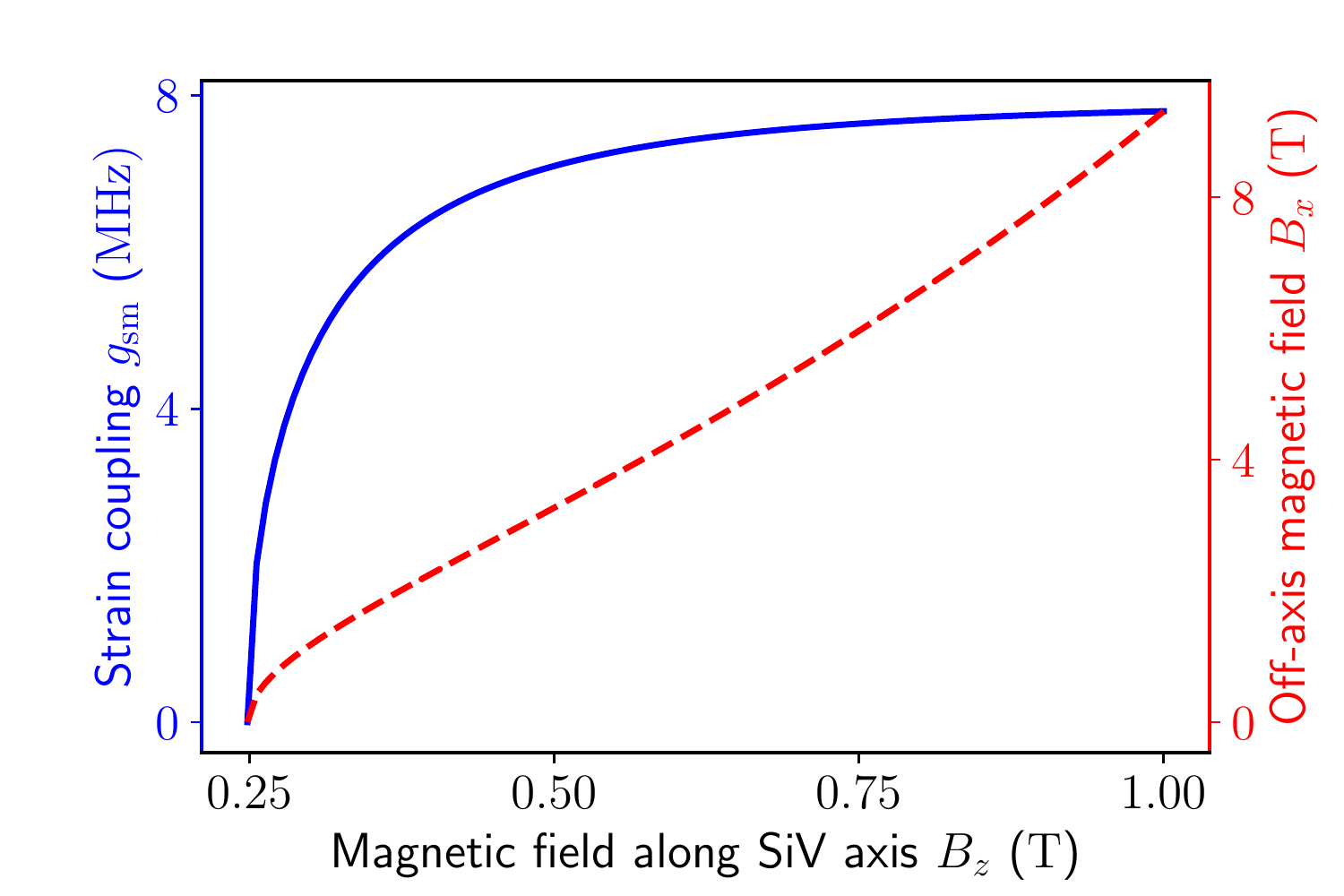}
	\caption{
		Strain coupling $\gspin$ of the effective two-level system formed by the states $\ket{e_- \downarrow}'$ and $\ket{e_+ \uparrow}'$ as a function of the magnetic field $B_z$ parallel to the SiV axis (solid blue curve). 
		The level splitting of the two-level system is kept constant, $E_{+\uparrow} - E_{- \downarrow} = \omspin$, by tuning the off-axis magnetic field $B_x$ (dashed red curve), see Eq.~\eqref{eqn:SM:SiVTuningBx}.
		Parameters are $\omspin = \Valueomegamech$, $\lambda_\mathrm{SO} = \ValuelambdaSO$, $\gamma_\mathrm{L} = \ValuegammaL$, $\gamma_\mathrm{S} = \ValuegammaS$, $\varepsilon_{E_{gx}} = \ValueepsilonEgx$, and $\varepsilon_{E_{gy}} = \ValueepsilonEgy$. 
	}
	\label{fig:SiVStrainCoupling}
\end{figure}

To estimate the strain coupling, we performed \textsc{COMSOL} simulations of a representative diamond OMC structure with the displacement chosen such that the total strain energy was equal to half the zero-point energy of the mechanical mode. 
The SiV defect was chosen to be centered vertically and laterally, and to be positioned half-way between the central hole and the adjacent hole along the long axis of the OMC. 
We considered two different SiV orientations in the diamond lattice and found the strain tensor to be of the order of $10^{-9}$, giving rise to  $\varepsilon_{E_{gx}} = \ValueepsilonEgx$, and $\varepsilon_{E_{gy}} = \ValueepsilonEgy$. 
The field-dependent strain coupling strength is then given by $\gspin = \abs{c_1 (\varepsilon_{E_{gx}} + i \varepsilon_{E_{gy}})} = \mathcal{O}(\mathrm{MHz})$ and is shown in Fig.~\ref{fig:SiVStrainCoupling}.

\subsection{Coherence properties of SiV defects}
\label{sec:SM:SiVCoherence}
Above temperatures of $\approx 1\,\mathrm{K}$, the coherence time of a negatively charged SiV center is limited by excitation processes from the lower doublet $\{ \ket{e_{g-} \downarrow}, \ket{e_{g+} \uparrow}\}$ to the upper doublet $\{ \ket{e_{g+} \downarrow}, \ket{e_{g-} \uparrow} \}$ of the orbital ground-state manifold. 
These two doublets are separated by $\Delta_\mathrm{gs} = \lambda_\mathrm{SO} \approx 50\,\mathrm{GHz}$ due to the spin-orbit interaction, such that thermally populated phonons in the diamond lattice can induce transitions \cite{Meesala2018}.  
While early studies of SiV centers were limited by this process to very short coherence times of $\approx 100\,\mathrm{ns}$ at $4\,\mathrm{K}$, several strategies to improve the decoherence time (which are compatible with our setup) have recently been proposed or even demonstrated:
\begin{itemize}
	\item Sukachev \emph{et al.} showed that cooling the setup in a dilution refrigerator to temperatures of about $100\,\mathrm{mK}$ improves the coherence time by five orders of magnitude to about $13\,\mathrm{ms}$ \cite{Sukachev2017}, and also suppress thermal occupation of the mechanical mode (since $\nth \ll 1$ for a several-GHz mechanical mode).
		This work as well as others \cite{Pingault2017} also demonstrated optically-detected magnetic resonance and Ramsey measurements, such that coherent control of SiV defects and quantum sensing with them is technically feasible. 
	\item Instead of cooling the setup, one can in principle engineer the phonon band structure of the diamond host material by designing it to be a phononic crystal with a band gap around $50\,\mathrm{GHz}$ \cite{Chia2021}. 
		The optomechanical crystal considered in our work is a special case of such a phononic crystal, which is engineered to serve simultaneously as a photonic crystal.
		To improve the SiV coherence time, the band structure of the OMC can be engineered to suppress $50\,\mathrm{GHz}$ phonons, or the OMC can be surrounded by a specifically designed phononic shield serving the same purpose. 
	\item Meesala \emph{et al.} showed that the splitting $\Delta_\mathrm{gs}$ can also be increased by applying static strain, and they demonstrated an increase from $\Delta_\mathrm{gs} = \lambda_\mathrm{SO} \approx 50\,\mathrm{GHz}$ to $\Delta_\mathrm{gs} \approx 400\,\mathrm{GHz}$ \cite{Meesala2018}.
		Another potential strategy (that requires further investigation) could thus be to apply a combination of dc strain, which increases $\Delta_\mathrm{gs}$ and improves the coherence time, and additional ac strain, which stems from the vibrations of the mechanical mode and generates the desired spin-mechanical interaction.
\end{itemize}

Having eliminated the typical source of dephasing, one may worry that the large value of the strain coupling $\gspin$ deteriorates the coherence properties of the SiV, too, because the two-level system of interest (with $\omspin \approx$ several GHz) may couple to other mechanical modes of the OMC. 
A one-dimensional OMC [as sketched in Fig.~\ref{fig:sketch}(b) of the main text] has only a pseudo-bandgap (which suppresses only localized modes with mirror symmetry perpendicular to the long axis of the OMC), such that resonant interactions with nonsymmetric and propagating phonon modes may lead to additional relaxation of the SiV center. 
However, these modes have a lower $Q$ factor, which reduces the spin cooperativity and leads to a suppression of undesired additional decay mechanisms.
Finally, dispersive interactions with off-resonant mechanical modes could cause additional dephasing of the SiV center due to thermal phonon-number fluctuations.
Even though dephasing is irrelevant for QND readout, it may limit the operation of the SiV as a quantum sensor. 
However, the associated dephasing rates will be highly suppressed because $\nth \ll 1$.

\subsection{Prospects for solid-state defects other than SiV}
\label{sec:SM:ProspectsOtherDefects}
So far, we analyzed the particular case of a negatively charged SiV defect embedded in a diamond OMC.
However, as we stress in the main text, the scheme is not restricted to SiV defects. 
The general criteria listed in Sec.~\ref{sec:SM:GeneralCriteria}, and the conditions~\eqref{eqn:SM:Criterion:Detuning} and~\eqref{eqn:SM:Criterion:CollectiveCooperativity} derived from the QND readout condition~\eqref{eqn:SM:Criterion:RelaxatinRates} can be satisfied if the single-spin strain coupling $\gspin$ or the collectively enhanced strain coupling $\sqrt{N} \gspin$ in an ensemble of $N$ spins are large.
Since strain susceptibilities have not yet been measured for many solid-state defects, providing an exhaustive list of all potentially relevant spin defects is impossible.
However, natural candidates for other defect systems are those that have a similar microscopic structure as NV or SiV defects in diamond, which translates into similar expected strain susceptibilities.

\subsubsection{SiV-like defects}
For SiV defects, the high strain susceptibility stems from the fact that the states $\{ \ket{e_{g-} \downarrow}, \ket{e_{g+} \uparrow}, \ket{e_{g+} \downarrow}, \ket{e_{g-} \uparrow} \}$ in the ground-state manifold are superpositions of different electronic orbitals (due to the strong spin-orbit interaction), which are relatively large compared to the carbon atoms that the defect replaces (since they are linear combinations of dangling carbon bonds adapted to the $D_{3d}$ symmetry). 
Strain deforms the crystal lattice, which changes the electronic orbitals and thus directly affects the SiV ground states.
These properties are expected to be present in other group-IV defect centers with the same $D_{3d}$ symmetry, e.g., GeV, SnV and PbV defects: 

\begin{itemize}
	\item A promising and well studied candidate is the negatively-charged tin-vacancy (SnV) center in diamond, which has a much larger spin-orbit splitting of $\approx 850\,\mathrm{GHz}$ (reducing phonon-induced dephasing) and a long coherence time of $\approx 0.3\,\mathrm{ms}$ \cite{Debroux2021}.
		This study also measured spin-strain coupling energies of $\approx 200\,\mathrm{GHz}$ in a sample with a static ``moderate-strain'' perturbation, which may hint at large spin-strain susceptibilities, comparable to those in SiV defects. 
		However, to the best of our knowledge, the spin-strain susceptibilities have not yet been measured. 
	\item Similarly, the neutral silicon-vacancy center (SiV0) has been shown to have long coherence times of $\approx 1\,\mathrm{ms}$ below $20\,\mathrm{K}$ \cite{Rose2018} and 
		to be sensitive to strain \cite{Green2019}, but strain susceptibilities have not yet been measured to the best of our knowledge. 
		Fluorescence readout has long been elusive for the SiV0 defect center and has only recently been achieved using bound exciton states \cite{Zhang2020}. 
		Our OMIT scheme may provide a convenient alternative readout technique for this spin defect. 
\end{itemize}

\subsubsection{NV-like defects}
For NV defects, all states in the ground-state manifold ${}^3 A_2$ have the same electronic wave function such that there is no differential change due to strain \cite{Maze2011}. 
The zero-field splitting $D \approx 2.87\,\mathrm{GHz}$ stems from the dominant spin-spin interaction, and the spin-strain coupling emerges only from a higher-order correction to the spin-spin interaction due to the spin-orbit interaction \cite{Lee2016}. 
Therefore, the strain susceptibilities are generally smaller than in SiV-like defects, but appreciable dispersive couplings $\chi$ may still be achievable in ensembles:

\begin{itemize}
	\item NV defects offer excellent coherence times (on the order of milliseconds) and have been demonstrated to be excellent quantum sensors for a variety of sensing targets, including magnetic fields, electric fields, and temperature. 				
		However, as a consequence of their electronic structure, they feature only a very small strain coupling of $\approx 20\,\mathrm{GHz}/\mathrm{strain}$ in the ground-state manifold \cite{Ovartchaiyapong2014,Teissier2014,Meesala2016}.
		This precludes single-spin readout, but appreciable dispersive shifts could still be achieved in a large ensemble ($N \approx 10^9$ defects).
		Since fluorescence readout of large NV ensembles is limited by low contrast and low photon collection efficiency due to the optical reset of the NV center \cite{Barry2020}, our OMIT scheme may still provide significant improvements for readout of large ensembles. 
	\item Divacancy defects in silicon carbide (3C-SiC and 4H-SiC) have been shown to have coherence times of $\approx 1\,\mathrm{ms}$ \cite{Seo2016,Christle2017} and strain susceptibilities of $\approx 2-4\, \mathrm{GHz}/\mathrm{strain}$ \cite{Falk2014}. 
		Again, these susceptibilities could yield appreciable dispersive shifts in large ensembles of defects.
		Also, SiC optomechanical microresonators with frequencies in the $\mathrm{GHz}$ range \cite{Lu2015} and SiC optomechanical crystals at $\mathrm{MHz}$ frequencies \cite{Lu2020} have already been demonstrated.
	\item Another solid-state defect with the same $C_{3v}$ symmetry is the negatively-charged monovacancy $V_\mathrm{Si}^-$ in 4H-SiC, which has coherence times of about $0.3\,\mathrm{ms}$ \cite{Soykal2017} and high strain susceptibilities \cite{Vasquez2020}.
\end{itemize}

\section{Quantum sensing using OMIT}
\subsection{Estimation error}
\label{sec:SM:QuantumSensing:EstimationError}

In this section, we take a step back and consider our proposed OMIT readout scheme in the broader context of quantum sensing. 
More specifically, we assume that a signal to be measured couples to the mechanical mode and causes a small shift $\varepsilon \lll \ommech,\Gmech,\kappa$ of the mechanical resonance frequency. 
The estimation error for such an infinitesimal signal is given by the fluctuations $\cerw{[\widehat{\delta \mathcal{I}}]^2}_\varepsilon$ of the homodyne current, referred back to the signal $\varepsilon$ by normalizing with the rate of change $\partial_\varepsilon \cerw{\hat{\mathcal{I}}}_\varepsilon$ with respect to variations in $\varepsilon$,
\begin{align}
	(\mathbf{\Delta}\varepsilon)^2(\tau) = \lim_{\varepsilon \to 0} \frac{\cerw{[\widehat{\delta\mathcal{I}}(\tau)]^2}_\varepsilon}{\cabs{\partial_\varepsilon \cerw{\hat{\mathcal{I}}(\tau)}_\varepsilon}^2} \fullstop
	\label{eqn:SM:EstimationError}
\end{align}
This estimation error is closely related to the SNR introduced in Eq.~\eqref{eqn:DefinitionSNR} of the main text, 
\begin{align}
	(\mathbf{\Delta}\varepsilon)^2(\tau) 
	= \lim_{\varepsilon \to 0} \frac{\frac{1}{2} \left( \cerw{[\widehat{\delta\mathcal{I}}]^2}_{+\varepsilon} + \cerw{[\widehat{\delta\mathcal{I}}]^2}_{-\varepsilon} \right)}{\abs{ \frac{1}{2 \varepsilon} \left( \cerw{\hat{\mathcal{I}}}_{+\varepsilon} - \cerw{\hat{\mathcal{I}}}_{-\varepsilon} \right) }^2}
	= \lim_{\varepsilon \to 0} \frac{2 \varepsilon^2}{\SNR^2(\tau)} \fullstop
\end{align}
Since we are considering the limit of an infinitesimal signal $\varepsilon \to 0$, the integration time will be much larger than the mechanical ringup time $1/\Gmech$ and we can ignore transient dynamics. 
The estimation error optimized over the homodyne detection angle $\varphi$ is 
\begin{align}
	(\mathbf{\Delta}\varepsilon)^2(\tau) = \frac{\left[ \Gmech^2 (1 + \Com)^2 + 4 \delta^2 \right] \left[ \Gmech^2 (1 + \Com)^2 + 8 \nth \Gmech^2 \Com + 4 \delta^2 \right]}{64 \Gmech^2 \Com^2 \tau \abs{a_\mathrm{pr,in}}^2 } \comma
\end{align}
which can be further optimized by choosing a resonant probe laser, $\delta = 0$, and by choosing the impedance-matching condition $\Com = 1$.
We thus find the optimal estimation error
\begin{align}
	(\mathbf{\Delta}\varepsilon)^2_\mathrm{opt}(\tau) 
	= \frac{\Gmech^2}{4 \abs{a_\mathrm{pr,in}}^2 \tau} (1 + 2 \nth) 
	= \frac{\Gmech}{4 \nmech^\mathrm{ss} \tau} (1 + 2 \nth) \comma
	\label{eqn:SM:optimizedSensitivity}
\end{align}
where we used the steady-state phonon number~\eqref{eqn:SM:nmechss} in the last step.

The OMIT measurement compares the small unknown frequency shift $\varepsilon$ to the width of the OMIT dip in the optical output power, which is given by $\Gmech (1 + \Com)$: 
For perfect impedance matching, $\Com = 1$, and no signal, $\varepsilon = 0$, the weak probe laser will resonantly drive the mechanical mode and the semiclassical amplitude of the optical output field will be zero. 
In the presence of small signal, $\varepsilon \neq 0$, the mechanical mode is slightly detuned from the probe laser and some photons will be emitted into the optical output field, yielding a finite semiclassical amplitude. 
The phase quadrature of the reflected light is linearly proportional to $\varepsilon$ if the condition $\varepsilon \lesssim \Gmech$ holds (which sets the dynamic range of this method). 
The mechanical decay rate (which is the smallest decay rate in the system) thus sets the ``ruler'' with which the unknown frequency shift $\varepsilon$ is compared. 
With this picture in mind, the impedance-matching condition marks the optimal trade-off between having no mechanical-to-optical conversion in the limit $\Com \to 0$, and an undesired ``stretching'' of the ``ruler'', $\Gmech \to \Gmech (1 + \Com) \to \infty$ in the limit $\Com \to \infty$.

\subsection{Comparison of different detection schemes}

It is instructive to compare our OMIT based detection with other methods to detect a small mechanical frequency shift.
In all schemes, the mechanical oscillator is driven at frequency $\ommech$. 
Its position operator,
\begin{align}
	\hat{x}(t) &= \sqrt{2} x_\mathrm{zpf} \left[ \xmech(t) \cos(\ommech t) + \pmech(t) \sin(\ommech t) \right] \comma 
\end{align}
can be decomposed into the mechanical cosine and sine quadratures
\begin{align}
    \xmech(t) &= \frac{1}{\sqrt{2}} (\mechmode^\dagger e^{-i \ommech t} + \mechmode e^{+i \ommech t}) \comma \\
    \pmech(t) &= \frac{i}{\sqrt{2}} (\mechmode^\dagger e^{-i \ommech t} - \mechmode e^{+i \ommech t}) \comma
\end{align}
where $x_\mathrm{zpf}$ denotes the mechanical zero-point fluctuations. 
We assume that the phase of the mechanical drive is chosen such that the oscillation of $\cerw{\hat{x}(t)}$ is purely sinusoidal in the absence of a signal, i.e., $\cerw{\xmech(t)} = 0$ for $\varepsilon = 0$.
In this case, the cosine quadrature $\xmech(t)$ is called the \emph{phase quadrature}. 
A change of the mechanical resonance frequency $\ommech \to \ommech + \varepsilon$ will change the relative phase between the mechanical drive and $\cerw{\hat{x}(t)}$, such that the phase quadrature $\cerw{\xmech(t)} \propto \varepsilon$ becomes nonzero and allows us to infer the frequency shift $\varepsilon$. 

\begin{figure}
	\centering
	\subfigure[]{
		\includegraphics[width=0.31\textwidth]{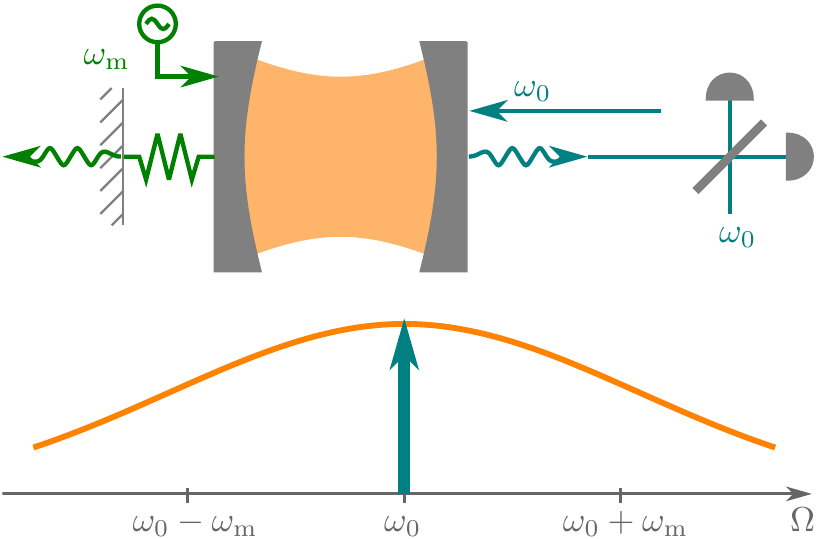}
	}
	\subfigure[]{
		\includegraphics[width=0.31\textwidth]{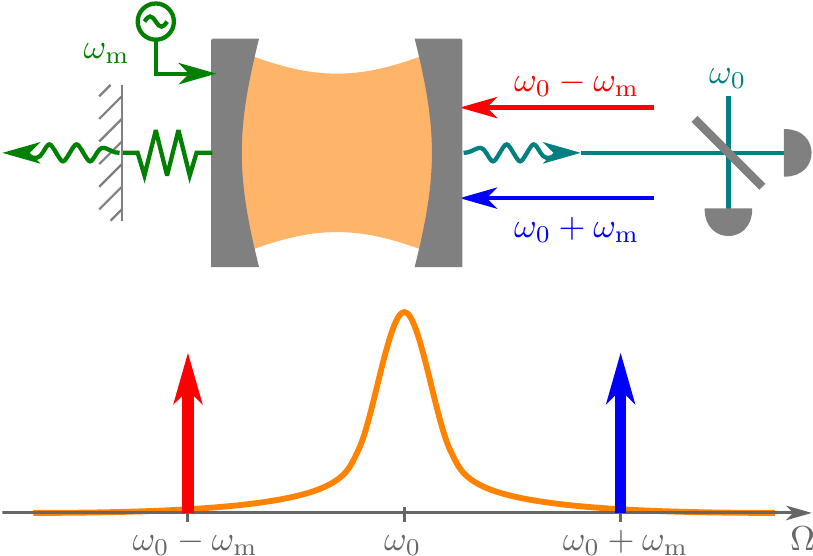}
	}
	\subfigure[]{
		\includegraphics[width=0.31\textwidth]{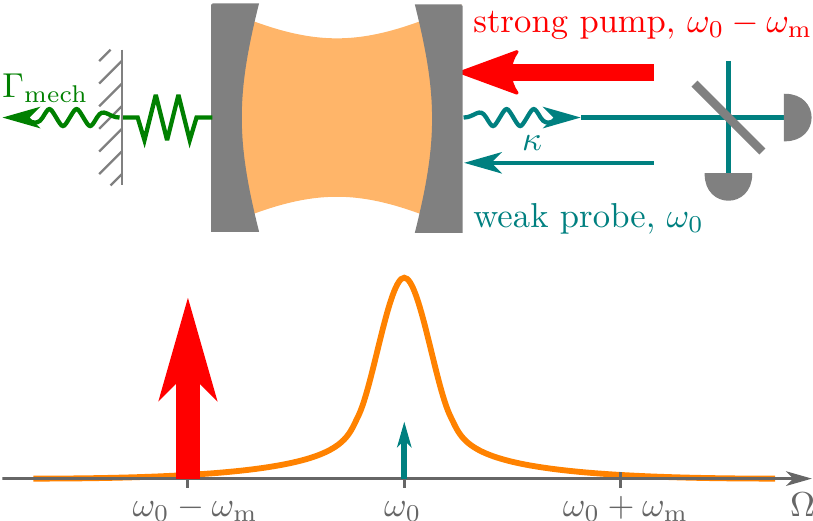}
	}
	\caption{
		Comparison of different methods to detect a small mechanical frequency shift $\ommech \to \ommech + \varepsilon$. 
		(a) Optomechanical position measurement, whose estimation error is limited to the standard quantum limit of position detection.
		(b) Backaction evading (BAE) measurement, which can surpass the standard quantum limit.
		(c) OMIT readout scheme, which reaches the same estimation error as a BAE measurement with experimentally less demanding requirements.
	}
	\label{fig:SM:Comparison_of_detection_schemes}
\end{figure}

\paragraph{Direct position measurement}
Perhaps the most obvious approach to measure $\cerw{\xmech(t)}$ is a direct position measurement, which is sketched in Fig.~\ref{fig:SM:Comparison_of_detection_schemes}(a) and has been analyzed in \cite{Clerk2010}. 
The oscillating mechanical position $\hat{x}(t)$ causes a parametric modulation of the optical cavity frequency, $\omopt(t) = \omopt [ 1 - (g_0/\omopt) \cerw{\hat{x}(t)}/x_\mathrm{zpf}]$, which leads to an oscillation of the phase of the light leaving the optical cavity. 
Depending on the local-oscillator phase, the homodyne detection measures one of the quadratures of the optical output field, which are given by
\begin{align}
    \xcavout(t) &= \frac{1}{\sqrt{2}} [\optfluc_\mathrm{out}^\dagger(t) e^{-i \omopt t} + \optfluc_\mathrm{out}(t) e^{+i \omopt t}] \comma \\
    \pcavout(t) &= \frac{i}{\sqrt{2}} [\optfluc_\mathrm{out}^\dagger(t) e^{-i \omopt t} - \optfluc_\mathrm{out}(t) e^{+i \omopt t}] \fullstop
\end{align}
For a suitably chosen phase of the optical input field, the $\pcavout$ quadrature of the optical output field is proportional to the mechanical position, 
\begin{align}
    \cerw{\xcavout(t)} &= 0 \comma &
    \cerw{\pcavout(t)} &= \sqrt{2 \Gmech \Com} \frac{\cerw{\hat{x}(t)}}{x_\mathrm{zpf}} \fullstop
\end{align}
The fundamental limitation of a direct position measurement stems from the fact that it amplifies the non-commuting mechanical quadratures $\xmech$ and $\pmech$ equally by a gain factor $\propto \sqrt{\Com}$. 
It can thus be understood as a phase-insensitive linear amplifier \cite{Caves1982}, 
\begin{align}
	\optfluc_\mathrm{out}[\ommech] = - \sqrt{\Com} \frac{2 i \Gmech}{\Gmech + 2 i \varepsilon} \mechmode_\mathrm{in}[\ommech] + \text{noise terms} \comma
	\label{eqn:SM:ModeAmplificationPOS}
\end{align}
where we switched to frequency space by defining the Fourier-transformed operator $\hat{O}[\omega] = \int_{-\infty}^\infty \d t\, \hat{O}(t) e^{i \Omega t}$. 
To ensure that the output modes $\optfluc_\mathrm{out}$ and $\optfluc_\mathrm{out}^{\dagger}$ have proper commutation relations, imprecision noise $\nmeas \geq 1/2$ has to be added during the amplification step, i.e., 
\begin{align}
	(\mathbf{\Delta}\varepsilon)^2(\tau) = \frac{\Gmech}{4 \nmech^\mathrm{ss} \tau} \left( 2 + 2 \nth \right) \fullstop
	\label{eqn:SM:SensitivityDirectPositionDetection}
\end{align}
Assuming thermal noise is negligible, $\nth \to 0$, the estimation error~\eqref{eqn:SM:SensitivityDirectPositionDetection} for a direct position measurement is thus at least a factor of $2$ larger than for our OMIT readout scheme, Eq.~\eqref{eqn:SM:optimizedSensitivity}. 
Note that the direct optomechanical position measurement does not require sideband resolution, i.e., it can be implemented in a system having $\kappa \gg \ommech$.
The additional factor of $2$ in the estimation error may be a reasonable price to pay for not having to operate in the regime $\kappa \ll \ommech$.

The fundamental limit $\nmeas = 1/2$ is called the standard quantum limit of position detection (SQL-PD) \cite{Braginsky1992,Clerk2010}. 
In principle, the SQL-PD also applies to the slope detection technique in atomic-force microscopy (AFM) \cite{Albrecht1991}, where one uses a detuned mechanical drive on the slope of the mechanical resonance curve, measures $\cerw{\hat{x}(t)}$, and infers $\varepsilon$ from the amplitude of oscillation. 
However, current setups are limited by thermal noise, i.e., $\nth \gg 1$. 
In this limit, Albrecht \emph{et al.} also showed that switching to FM detection, where the mechanical oscillator is driven into limit-cycle motion, does not improve the estimation error over slope detection (but improves the dynamic range) \cite{Albrecht1991}. 
The same result has been found for sensors using optomechanical limit cycles \cite{Guha2020}.

\paragraph{Backaction-evading measurement}

The SQL-PD can be surpassed by a backaction-evading (BAE) measurement of $\xmech(t)$ \cite{Braginsky1980,Clerk2008}, which is sketched in Fig.~\ref{fig:SM:Comparison_of_detection_schemes}(b).
In this scheme, two laser drives of equal amplitude are applied on the red and blue mechanical sideband.
Their relative phase is chosen such that the cavity resonance frequency depends (in a time-average way) only on $\xmech(t)$, i.e., the interaction term $\propto \optmode^\dagger \optmode \hat{x}(t)$ is replaced by a term of the form $ \optmode^\dagger \optmode [ \xmech(t) + \text{terms averaging to zero}]$. 
For a suitably chosen phase of the optical output input field, one now finds
\begin{align}
    \cerw{\xcavout(t)} &= 0 \comma &
    \cerw{\pcavout(t)} &= \sqrt{\Gmech \Com} \cerw{\xmech(t)} \fullstop
\end{align}
Unlike direct position detection, only the mechanical $\xmech(t)$ quadrature is amplified. 
The BAE scheme thus implements a phase-sensitive amplification scheme, and the equivalent of Eq.~\eqref{eqn:SM:ModeAmplificationPOS} (written in terms of quadrature operators) takes the form,
\begin{subequations}%
\begin{align}%
	\xcavout[0] &= 0 + \text{noise terms} \comma 
	\label{eqn:SM:ModeAmplificationBAE1} \\
	\pcavout[0] &= \sqrt{\Gmech \Com} \xmech[0] + \text{noise terms} \fullstop
	\label{eqn:SM:ModeAmplificationBAE2}
\end{align}%
\label{eqn:SM:ModeAmplificationBAE}%
\end{subequations}%
Also in the BAE scheme, noise must be added to ensure that the optical output modes have proper canonical commutation relations.
However, the amount of added noise is independent of $\Com$ and becomes irrelevant in the limit of large gain. 
The estimation error on changes $\varepsilon$ of the mechanical resonance frequency for such a BAE measurement is
\begin{align}
	(\mathbf{\Delta}\varepsilon)^2(\tau) = \frac{\Gmech}{4 \nmech^\mathrm{ss} \tau} \frac{1 + 8 \Com (1 + 2 \nth)}{8 \Com} 
	\stackrel{\Com \to \infty}{\longrightarrow} \frac{\Gmech}{4 \nmech^\mathrm{ss} \tau} (1 + 2 \nth) \fullstop
	\label{eqn:SM:SensitivityBAEPositionDetection}
\end{align}
Hence, in the limit of a large optomechanical cooperativity (i.e., large gain), the optomechanical BAE measurement achieves the same estimation error as our OMIT detection scheme. 
However, our OMIT detection scheme has an experimentally much more forgiving condition on the required cooperativity, $\Com = 1$.
Whereas OMIT-type experiments are routinely used for characterization of optomechanical setups, BAE measurements \cite{Hertzberg2010,Suh2014,Lecocq2015,OckeloenKorppi2016,Shomroni2019} are still much more challenging.

\paragraph{OMIT measurement}

For completeness, we also give the equivalent of Eqs.~\eqref{eqn:SM:ModeAmplificationPOS} and \eqref{eqn:SM:ModeAmplificationBAE} for the OMIT detection scheme: 
\begin{align}
	\optfluc_\mathrm{out}[\ommech] = \frac{\Gmech (\Com - 1) - 2 i \varepsilon}{\Gmech (\Com + 1) + 2 i \varepsilon} \optfluc_\mathrm{in}[\ommech] + \text{noise terms}
\end{align}
Similar to a direct position measurement, OMIT implements a phase-insensitive amplification scheme, but the gain factor never exceeds unity. 
Therefore, no imprecision noise has to be added to preserve the commutation relations of the output modes and $\nmeas = 0$ is possible, similar to a BAE measurement in the limit of large gain, $\Com \to \infty$.
Note that the absence of amplification in the OMIT scheme is not a drawback:
Unlike optomechanical force sensing, where a small unknown force leads to a tiny change of the mechanical \emph{position} that needs to be amplified for detection, we want to detect a small change of the mechanical resonance frequency, which manifests itself in a change of the \emph{phase} of oscillation. 
Such a phase change can be probed by driving the mechanical oscillator strongly, which ensures a large output signal and eliminates the need for amplification.

Note that both optomechanical position measurements and BAE detection require a \emph{mechanical} drive whose phase needs to be carefully tuned with respect to the optical local-oscillator used in the homodyne detection setup. 
In contrast, as shown in Fig.~\ref{fig:SM:Comparison_of_detection_schemes}(c), our OMIT detection scheme uses the \emph{optical} probe laser as a resonant mechanical drive. 
Both the weak probe laser and the local-oscillator signal can thus be derived from the same source, which enables a convenient all-optical measurement of the phase of the reflected light. 
Finally, it is interesting to note that Eq.~\eqref{eqn:SM:optimizedSensitivity} is only a factor of $(1 + \Com)^2/\Com = 4$ larger than the estimation error one could obtain in a hypothetical direct mechanical homodyne detection of the phonons dissipated into the substrate (green wiggly arrows in Fig.~\ref{fig:SM:Comparison_of_detection_schemes}).

\subsection{Impact of imperfect homodyne detection}
\label{sec:SM:QuantumSensing:ImperfectHD}

Experimentally, the estimation error of the OMIT measurement will be limited by the efficiency $\eta$ of the homodyne detection. 
Imperfect detection can originate both from the finite efficiency of the photon detectors as well as from scattering and absorption losses on the way to the detector. 
All these imperfections can be modeled by assuming an additional beamsplitter with transmittivity $\eta < 1$ between the optomechanical system and the 
homodyne detection, which mixes the output mode $\optfluc_\mathrm{out}(t)$ with vacuum noise $\hat{\xi}_\mathrm{imp}(t)$ and therefore discards a fraction $\sqrt{1 - \eta}$ of the optical output field. 
The two output modes of the beamsplitter are  
\begin{align}
	\hat{o}_1(t) &= \sqrt{\eta} \optfluc_\mathrm{out}(t) + i \sqrt{1 - \eta} \hat{\xi}_\mathrm{imp}(t) \comma \\
	\hat{o}_2(t) &= -i \sqrt{1 - \eta} \optfluc_\mathrm{out}(t) - \sqrt{\eta} \hat{\xi}_\mathrm{imp}(t) \comma
\end{align}
where $\hat{\xi}_\mathrm{imp}(t)$ is zero-temperature Gaussian white noise, $\cerw{\hat{\xi}_\mathrm{imp}(t) \hat{\xi}_\mathrm{imp}^\dagger(t')} = \delta(t - t')$ and $\cerw{\hat{\xi}_\mathrm{imp}^\dagger(t) \hat{\xi}_\mathrm{imp}(t')} = 0$.
The output mode $\hat{o}_1(t)$ is measured by the homodyne detection setup, such that Eq.~\eqref{eqn:SM:IntegratedHomodyneCurrentOperator} is replaced by the observable
\begin{align}
	\hat{\I}_\mathrm{imp}(\tau) 
	&= \sqrt{\kappa} \int_0^\tau \d t\, \left[ e^{i \varphi} e^{-i \omprrot t} \hat{o}_1^\dagger(t) + e^{-i \varphi} e^{i \omprrot t} \hat{o}_1(t) \right] \nonumber \\
	&= \sqrt{\eta} \hat{\I}(\tau) - i \sqrt{1 - \eta} \sqrt{\kappa} \int_0^\tau \d t\, \left[ e^{i \varphi} e^{-i \omprrot t} \hat{\xi}_\mathrm{imp}^\dagger(t) - e^{-i \varphi} e^{i \omprrot t} \hat{\xi}_\mathrm{imp}(t) \right] \fullstop
\end{align} 
The homodyne detection signal is thus rescaled by $\sqrt{\eta}$ whereas the fluctuations are partially replaced by integrated vacuum noise,
\begin{align}
	\cerw{\hat{\I}_\mathrm{imp}(\tau)}_\varepsilon &= \sqrt{\eta} \cerw{\hat{\I}(\tau)}_\varepsilon \comma \\
	\cerw{[\widehat{\delta\I}_\mathrm{imp}]^2}_\varepsilon &= \eta \cerw{[\widehat{\delta\I}]^2}_\varepsilon + (1 - \eta) \tau \fullstop
\end{align}
The estimation error~\eqref{eqn:SM:EstimationError} is now given by
\begin{align}
	(\mathbf{\Delta}\varepsilon)^2_\mathrm{imp}(\tau) 
	= \lim_{\varepsilon \to 0} \frac{\cerw{[\widehat{\delta\I}_\mathrm{imp}]^2}_\varepsilon}{\cabs{\partial_\varepsilon \cerw{\hat{\I}_\mathrm{imp}(\tau)}_\varepsilon}^2} 
	&= (\mathbf{\Delta}\varepsilon)^2(\tau) + \frac{1 - \eta}{\eta} \frac{\left[ \Gmech^2 (1 + \Com)^2 + 4 \delta^2 \right]^2}{64 \Gmech^2 \Com^2 \tau \abs{a_\mathrm{pr,in}}^2} \fullstop
\end{align}
The additional term due to imperfect homodyne detection does not change the optimal parameters $\delta = 0$ and $\Com = 1$ found in Sec.~\ref{sec:SM:QuantumSensing:EstimationError}, such that we obtain the minimum estimation error
\begin{align}
	(\mathbf{\Delta}\varepsilon)^2_\mathrm{imp,opt}(\tau)
	= (\mathbf{\Delta}\varepsilon)^2_\mathrm{opt}(\tau) + \frac{1 - \eta}{\eta} \frac{\Gmech}{4 \nmech^\mathrm{ss} \tau}  \comma
\end{align}
where we used the definition~\eqref{eqn:SM:nmechss} for $\nmech^\mathrm{ss}$ to simplify the expression. 
Comparing this result with Eq.~\eqref{eqn:SM:optimizedSensitivity}, we see that we can rewrite the imperfect-detection term as an equivalent amount of thermal phonons, 
\begin{align}
	\ndet = \frac{1 - \eta}{2 \eta} \comma
\end{align}
such that the optimized estimation error takes the form
\begin{align}
	(\mathbf{\Delta}\varepsilon)_\mathrm{opt}^2(\tau) = \frac{\Gmech}{4 \nmech^\mathrm{ss} \tau} \left( 1 + 2 \nth + 2 \ndet \right) \fullstop
	\label{eqn:SM:SensitivityOMITImperfect}
\end{align}
Figure~\ref{fig:sensing} of the main text compares Eq.~\eqref{eqn:SM:SensitivityOMITImperfect} with the corresponding sensitivities for direct position detection and a backaction evading measurement. 
OMIT readout surpasses the SQL of position detection for $\eta \geq 0.5$. 
Since homodyne detection efficiencies $\eta \gtrsim 0.7$ have already been demonstrated experimentally \cite{Purdy2013}, sensitivities beyond the SQL of position detection are feasible with state-of-the-art technology.

\section{Difference to optical superconducting qubit readout using microwave-to-optical transduction}
\label{sec:SM:DifferenceOpticalSCReadout}

In this section, we comment on the differences between our OMIT-based spin readout and a recently demonstrated optical readout of a superconducting qubit using microwave-to-optical transduction \cite{Delaney2022}. 
In this impressive experiment, Delaney \emph{et al.} couple a superconducting transmon qubit dispersively to a microwave cavity, which is optomechanically coupled to a mechanical mode of a silicon-nitride membrane. 
In addition, the mechanical mode is optomechanically coupled to an optical cavity, such that microwave-to-optical transduction can be achieved by applying simultaneous microwave and optical drives that are red-detuned from the respective cavity resonance frequencies by a mechanical frequency. 
The qubit is read out by sending a microwave pulse into the microwave cavity and transducing the reflected pulse (whose phase quadrature contains information on the qubit's state) into an optical output pulse. 
This setup requires simultaneous optimization of the microwave-to-mechanical and optical-to-mechanical coupling, which is technically challenging but can be achieved by spatially separating the interaction regions with the microwave and optical modes on the membrane. 

In our scheme, we eliminated the intermediary microwave mode by directly coupling the spins to the mechanical mode. 
This reduces experimental complexity since simultaneous optimization of two different optomechanical couplings is no longer required. 
Moreover, Delaney \emph{et al.} probe the qubit by applying a microwave pulse, which would correspond to a mechanical drive in our setup. 
We do not need this separate mechanical drive since the optical probe laser acts as a mechanical drive, which gives rise to a convenient \emph{all-optical} readout protocol. 
Finally, unlike in superconducting qubits, the microwave coupling of single solid-state spins is tiny, which precludes using Delaney \emph{et al.}'s readout scheme in a solid-state platform. 

\end{document}